\documentclass[11pt,a4paper]{article}
\usepackage{jheppub}
\usepackage[utf8]{inputenc}
\usepackage{graphicx,fancybox,float,comment}
\usepackage{units}
\usepackage{subcaption}
\usepackage{multirow,array,arydshln}
\usepackage{xcolor}
\usepackage{amsmath,amssymb}
\usepackage{cleveref}
\usepackage{yfonts}
\usepackage{tensor}
\usepackage{paralist}
\usepackage{braket}
\usepackage{tensor,mathrsfs}

\newcommand{\be}{\begin{eqnarray}}
\newcommand{\ee}{\end{eqnarray}}
\usepackage{ulem}
\usepackage{dsfont}
\DeclareMathAlphabet{\mathpzc}{OT1}{pzc}{m}{it}

\title{
A New Paradigm for Topological or Rotational Non-Abelian Gauge Fields from Einstein-Skyrme Holography
}
\author{Casey Cartwright, Benjamin Harms, Matthias Kaminski }

\affiliation{Department of Physics and Astronomy, University of Alabama,\\ Tuscaloosa, AL 35487, USA}

\emailAdd{cccartwright@crimson.ua.edu}
\emailAdd{bharms@ua.edu}
\emailAdd{mski@ua.edu}

\begin{document}
\newcommand{\eref}[1]{eq.\ (\ref{#1})}
\newcommand{\exd}{\mathrm{d}}
\newcommand{\reals}{\mathbb{R}}
\newcommand{\mB}{\mathcal{B}}
\newcommand{\pd}{\partial}
\newcommand{\s}{\sigma}
\newcommand{\ads}{$AdS_5$ }
\newcommand{\mbf}[1]{\mathbf{#1}}
\definecolor{amaranth}{rgb}{0.9, 0.17, 0.31}
\newcommand{\BH}[1]{\textcolor{orange}{ [Ben:\,#1]}}
\newcommand{\CC}[1]{\textcolor{blue}{ [Casey:\,#1]}}
\newcommand{\vev}[1]{\braket{#1}}
\newcommand\MK[1]{\textcolor{magenta}{ [Matthias:\,#1]}}
\newcommand{\tr}{\text{Tr}}
\newcommand{\kb}{\bar{\kappa}}

\abstract{
We report analytically known states at non-zero temperature which may serve as a powerful tool to reveal common  topological and thermodynamic properties of systems ranging from the QCD phase diagram to topological phase transitions in condensed matter materials. 
In the holographically dual gravity theory, these are analytic solutions to a five-dimensional non-linear-sigma (Skyrme) model dynamically coupled to Einstein gravity. 
This theory is shown to be holographically dual to $\mathcal{N}=4$ Super-Yang-Mills theory coupled to an $SU(2)$-current. 
%
All solutions are fully backreacted asymptotically Anti-de Sitter~(AdS) black branes or holes. 
One family of global $AdS$ black hole solutions contains non-Abelian gauge field configurations with positive integer Chern numbers and finite energy density. 
Larger Chern numbers  
increase the Hawking-Page transition temperature.
In the holographically dual field theory this indicates a significant effect on the deconfinement phase transition. 
Black holes with one Hawking temperature can have distinct Chern numbers, potentially enabling topological transitions. 
A second family of analytic solutions, rotating black branes, is also provided. These rotating solutions induce states with propagating charge density waves in the dual field theory.    
We compute 
the Hawking temperature, entropy density, angular velocity and free energy for these black
holes/branes. 
These correspond to thermodynamic data in the dual field theory. For these states the energy-momentum tensor, (non-)conserved current, and topological charge are interpreted.  
}

\maketitle

\section{Introduction} 
Topology has garnered much attention in a diverse number of sub-fields of physics. Topological solitons~\cite{Manton:2004tk} for instance have been the subject of intense study. 
%
Historically prominent examples include the
magnetic monopole in electromagnetism, the Skyrmion as a model for baryons in quantum chromodynamics~(QCD), screw dislocations in a crystalline lattice as well as the magnetic Skyrmion in condensed matter physics, and cosmic strings in cosmology. 

%
Furthermore, a better understanding of topological phases and transitions between these phases is required from both, a condensed matter perspective as well as from a high energy perspective. 
Topological phases of matter~\cite{Wen:2017} and symmetry protected topological phases~\cite{Senthil:2015} as well as transitions between such topological phases are a driving force in condensed matter theory and experiment. 
Gluon gauge fields are present in the strongly-coupled quark-gluon-plasma which is currently created at the Relativistic Heavy-Ion Collider (RHIC). These gluon field configurations are expected to undergo topological transitions (sphaleron processes)~\cite{Kharzeev:2004ey,Abelev:2009ac,Abelev:2009ad}. Such transitions can create states with chiral imbalance in which transport effects occur that are induced by chiral anomalies~\cite{Son:2009tf}. 
%
In order to better understand these topological processes in the elusive strongly-coupled regime of QCD at finite temperature, chemical potential, and within strong electromagnetic fields, topological solutions are in high demand. 
Holographic~\cite{Maldacena:1997re} models are able to provide such topological solutions as we show explicitly in this work. 

In addition, rotating fluids have also recently come into the focus of experiments in the mentioned research areas.  
%
Currently, RHIC creates quark-gluon-plasma states of matter which constitute the ``most perfect'' and ``most vortical'' fluids~\cite{STAR:2017ckg} realized in an experiment. 
%
Rotating systems are also of interest to condensed matter and ultracold gas experiments which, for example, are currently being carried out in helium-4 Bose-Einstein condensates~\cite{Bourne_2006}, or in liquid metals with nonzero vorticity~\cite{liquidMetalSpinHydroTakahashi2016}.

In this work we address the need for theoretical understanding of the aforementioned experiments. We provide a tool to study topological or rotating gauge field configurations in strongly coupled field theories through the use of the holographic principle~\cite{tHooft:1993dmi,Susskind:1994vu}. 
%
We employ a particular instance of holography, namely the gauge/gravity correspondence~\cite{Maldacena:1997re}. Our calculations and results within Einstein-Skyrme gravity map to results within the holographically dual strongly-coupled boundary field theory. 
%
Many solutions to gravity with asymptotically $AdS$-boundary conditions are known.  Examples include the {\it $AdS$ black hole}~\cite{Banados:1992gq} and the {\it $AdS$ soliton}~\cite{Horowitz:1998ha}. These two solutions are particularly illuminating examples for our purposes due to their topology. The $AdS$ black hole in Einstein gravity by conjecture corresponds to a thermal plasma in the dual boundary theory, which happens to be $\mathcal{N}=4$ Super-Yang-Mills theory. One may think of this as a model for a deconfined phase~\cite{Witten:1998zw} such as the quark-gluon plasma phase of QCD at high temperatures.\footnote{Note that the $AdS$ black brane with flat horizon topology does not display any deconfinement-like phase transition. However, a black hole with compact horizon at an infinite number of colors, $N$, does feature such a transition between a low and a high temperature phase~\cite{Witten:1998zw}.} 
%
In contrast to that, the $AdS$ soliton~\cite{Horowitz:1998ha} can be thought of as modeling a confining phase. This solution has one single compact coordinate direction which is spatial and which ends at a finite value of the $AdS$ radial coordinate (perpendicular to the conformal boundary). In this sense, this solution excises a part of space-time, as does the $AdS$ black hole. The soliton has topology $\mathbb{R}^{3,1}\times S^1$, while the black hole has $\mathbb{R}^{1,1}\times S^3$. 
%
These two examples realize two distinct five-dimensional bulk topologies, leading to a gravitational view on thermal field theories and confinement in the dual field theories but are devoid of topologically interesting gauge field or matter configurations.  

Including gravitating matter leads to a much richer set of possible phase structures in the dual theory. 
%
Solutions of this type may or may not have horizons, and have been classified as versions of the $AdS$ black hole or the  $AdS$ soliton with hair, respectively~\cite{Henneaux:2002wm,Banados:2005hm,Brihaye:2013tra,Anabalon:2016izw}. 
%
In \cite{Harms:2016pow,Harms:2017yko} 
new types of such asymptotically $AdS$ solutions in three space-time dimensions were found for rotating black holes. 
%
Matter is introduced as a non-linear $\sigma$-model contribution to the action, and two types of solutions were found, namely self-gravitating topological solitons and black holes with hair. 
It is important to note that Skyrmions appear in a distinct way  in~\cite{Harms:2016pow,Harms:2017yko} compared to previous approaches by Son/Stephanov and Sakai/Sugimoto which discuss Skyrmions on the field theory side of the gauge/gravity correspondence~\cite{Son:2003et,Sakai:2004cn}. 

%
In~\cite{Son:2003et,Sakai:2004cn} Skyrmions defined on the field theory side of the gauge/gravity correspondence are interpreted as baryons~\cite{Manton:2004tk} in the field theory, see table~\ref{tab:gaugeGravitySkyrmions}. 
The Sakai-Sugimoto model is very successful in modeling properties of QCD and has received much attention. For a review and literature references see~\cite{Rebhan:2014rxa}, and the more recent~\cite{Bigazzi:2020phm}. 
In~\cite{Sakai:2004cn} the gravitational object which is dual to such Skyrmions is a $D4$-brane wrapped around an $S^4$ constructed as a soliton in the world-volume gauge theory of the probe $D8$-branes~\cite{Sakai:2004cn,Hata:2007mb}. 
%
This is a top-down construction with parameters fixed by string theory. Gravity is dynamical in this setting. 
%
We will discuss realizations of baryons based on the Sakai/Sugimoto-model~\cite{Sakai:2004cn} in section~\ref{sec:holoSkyrmionsMerons}.

In~\cite{Son:2003et}, on the other hand, the gravity object is a five-dimensional instanton in a curved but non-dynamical metric background. 
%
More precisely, the authors consider a (3+1)-dimensional open moose theory, which is a non-linear sigma model with $SU(2)$-valued fields interacting through a gauge field. 
%
This theory is then lifted to five dimensions. Put in the context of a gauge/gravity correspondence, this is a bottom-up model.\footnote{
In addition to the models we explicitly compare to in this work, numerous other holographic models of topological phases have been constructed for example in~\cite{Karch:2010mn,Ryu:2010fe,Ryu:2010hc,HoyosBadajoz:2010ac}, see also references therein.}

Our approach in this paper is different from both these setups.\footnote{\label{foo:modelComparison} Our work may loosely be viewed as a ``hybrid'' extension of~\cite{Son:2003et} and~\cite{Sakai:2004cn}. Our setup is top-down embedded in string theory and treats gravity dynamically in the bulk, vaguely resembling the idea of~\cite{Sakai:2004cn}. We combine this with a topological gauge field configuration in a non-linear sigma model, resembling the spirit of~\cite{Son:2003et}. Our model is mathematically distinct, and on a qualitative level it combines the strong suits of the two previous models, eliminating many drawbacks.} 
%
We follow~\cite{Harms:2016pow,Harms:2017yko} and dynamically couple Skyrmions to gravity in the $AdS_5$ bulk, i.e.~our Skyrmions live on the gravity side of the gauge/gravity correspondence. 
%
A comparison of our setup to the previous setups is provided in table~\ref{tab:gaugeGravitySkyrmions}. 
A striking feature of our work is a consistent top-down embedding of $AdS$-bulk Skyrmion matter fields into holography. 
This is facilitated by an invertible map from the Einstein-Skyrme action to the Einstein-Yang-Mills action~\cite{Ipinza:2020xgc}, which we discuss in section~\ref{sec:merons} and which we show explicitly for our solutions in section~\ref{sec:blackBraneSolutions}.  
The Einstein-Yang-Mills action we work with is a consistent truncation of the bosonic part of minimal gauged type IIB supergravity in five dimensions, which is known to be dual to $\mathcal{N}=4$ Super-Yang-Mills~(SYM) theory~\cite{Buchel:2006gb,Gauntlett:2006ai,Gauntlett:2007ma}. The Yang-Mills $SU(4)$ gauge symmetry on the gravity side corresponds to the global $SU(4)$ R-symmetry of $\mathcal{N}=4$ SYM theory. It is also this map which allows us to interpret the effect of the bulk Skyrmion matter through the proxy of bulk meron Yang-Mills fields in section~\ref{sec:holoSkyrmionsMerons}.\footnote{\label{foo:ChernSimonsNeglected} 
In order to maintain the full supersymmetry of $\mathcal{N}=4$ SYM theory, we would have to add a Chern-Simons term to our action with a particular value of the Chern-Simons coupling. For simplicity we here explore solutions excluding this term first. Including this term is a straightforward generalization left for future investigation.} 
%
This offers a means to construct topologically non-trivial gauge field configurations not only in the bulk $AdS$ space-time but also in the boundary field theory, $\mathcal{N}=4$ SYM theory, and to holographically interpret them.

We direct the busy expert reader to the discussion section~\ref{sec:Discussion} for an overview and the main outcomes of this work, as well as future directions. Meanwhile the patient reader may want to begin with the following section~\ref{sec:SkyrmionsMeronsHolography} where the map between Einstein-Skyrme and Einstein-Yang-Mills theory is established. 
We first describe the Skyrmion matter content of our theory in section~\ref{sec:Skyrmions} and discuss the mapping of this theory to an Einstein-Yang-Mills theory in section~\ref{sec:merons}, more specifically to a meron subsector of Einstein-Yang-Mills theory. 
We then provide a review of the topological quantities which characterize our theory of meron Yang-Mills fields in section~\ref{sec:topologicalCharges}. 

Our main result is a new approach for constructing and holographically interpreting models of strongly-coupled phases of matter with topological gauge field configurations. 
Concretely, we provide three new analytic Einstein-Skymion/Einstein-Yang-Mills solutions in section~\ref{sec:blackBraneSolutions}. First, in section~\ref{sec:staticBBs}, we analyze a 
\textit{static Skyrme-$AdS_5$ black brane} with Skyrmion hair and obtain analytic solutions for the metric tensor elements which are very similar to those of a Reissner-Nordstr\"om black hole. As the second analytic solution, we then describe a \textit{rotating Skyrme-$AdS_5$ black brane} with Skyrmion hair which is obtained by utilizing a constraint previously found by one of the current authors~\cite{Harms:2016pow,Harms:2017yko} in section~\ref{sec:rotatingBBs}. Our third example, in section~\ref{sec:topologicalMeronSolutions}, is the asymptotically global ($\mathbb{R}^{1,1}\times S^3$) 
\textit{topological Skyrme-$AdS_5$ black hole} which carries non-trivial topological charge. 
In section~\ref{sec:holoSkyrmionsMerons} we discuss the holographic interpretation of our solutions including the extraction of the dual currents and energy-momentum tensor associated with the solutions presented in section~\ref{sec:blackBraneSolutions}. That section is concluded with a computation of the topological charge of the presented solutions. 

The results of our work form the foundation of a more generalized treatment of topological phases of matter at strong coupling. 
Our technique also allows an analytic study of such transitions. As a proof of principle, in section~\ref{sec:topologicalMeronSolutions}, we compute the free energy and study the topologically distinct phases of the boundary gauge field configuration dual to our analytic {topological Skyrme-$AdS_5$ black hole solution}, ensued by a discussion of its possible topological and phase transitions, extended in section~\ref{sec:comparisoToSS}. 
Gauge fields in our analysis are in principle not restricted to stationary configurations and hence can characterize a broader range of the dynamics associated with topological phases including their transitions. 
%
\begin{table}
\begin{center}
  \begin{tabular}{ l | c | c | r }
     & Son/Stephanov  & Sakai/Sugimoto & this paper \\ \hline\hline
    4+1d gravity dual & instanton   & probe $D4$-brane  & gravitational Skyrmion \\ 
    in $AdS_5$        & in external metric & in string theory & in Einstein-Skyrme theory  \\ \hdashline
    metric is:    & non-dynamical & dynamical & dynamical \\ \hdashline
   gauge field is:  & not backreacted & not backreacted & backreacted \\ \hline
    3+1d gauge theory &  Skyrmion & CFT Skyrmion & $\mathcal{N}=4$ Super-Yang-Mills  \\
    in flat $\mathbb{R}^{3,1}$ & = baryon & =  baryon  & coupled to (non-)conserved  \\
     &  &   &  $SU(2)$ current $J$ \\
     temperature:  &  $T=0$ & $T=0$  & $T\neq 0$ \\\hdashline
    topological charge: & baryon number & baryon number & winding/Chern number $\mathpzc{q}$ \\ \hline
  \end{tabular}
  \caption{\label{tab:gaugeGravitySkyrmions}
  Gravity Skyrmions in this paper versus field theory Skyrmions~\cite{Son:2003et,Sakai:2004cn} and the respective sides of the gauge/gravity correspondence. See main text and footnote~\ref{foo:modelComparison} for a more detailed comparison.
  }
\end{center}
\end{table}

\section{Skyrmions and merons}
\label{sec:SkyrmionsMeronsHolography}
%
\subsection{Skyrmions}
\label{sec:Skyrmions}
The Skyrme model was introduced in 1961~\cite{Skyrme:1961vq} as a non-linear model for pions. The fundamental field is an $SU(2)$ valued scalar $U(x,t)$. The Skyrmions in (3+1)-dimensional flat space-time have been discussed as descriptions of pions with their topological charge identified as a baryon number. 
Previous studies have investigated the Skyrme model coupled to gravity (Einstein-Skyrme model) in $(3+1)$-dimensional space-time and have shown the existence of solitons and black hole solutions with hair~\cite{Heusler:1991xx,Heusler:1993ci,Glendenning:1988qy,Piette:2007wd,Nelmes:2011zz,Luckock:1986tr,Bizon:1992gb,Kleihaus:1995vq,Tamaki:2001wca,Sawado:2004yq,Brihaye:2005an,Nielsen:2006gb,Duan:2007df,Doneva:2011gx,Gibbons:2010cr,Canfora:2013osa,Dvali:2016mur,Adam:2016vzf,Gudnason:2016kuu}. The quadratic term in the chiral fields present in these articles often led authors to numerical solutions of the field equations instead of analytic solutions for at least some of the fields.  
In contrast to previous studies, in our work we seek analytic, fully backreacted, asymptotically $AdS_5$ solutions to the Einstein-Skyrme model. 

The basic Skyrmion action in five dimensions is given by,
\be \label{eq:skyrmeAction}
S = \int d^5x \sqrt{-g}\left(\frac{R-2 \Lambda}{\kappa}+\mathcal{L}_m\,\right)  \label{eq:lagrange},
\ee
in which $R$ is the Ricci scalar, $\Lambda=-6/L^2$ in $AdS_5$ is the cosmological constant, $L$ is the $AdS_5$ radius, $\kappa\,=\,16\,\pi\,G$, $G$ is the five-dimensional gravitational constant, and $\mathcal{L}_m$ is the matter contribution to the Lagrangian density from the Skyrme field with coupling constant $\tilde{e}$. That matter contribution is
\begin{equation} 
\mathcal{L}_m\,=\,\frac{f_{\pi}}{16\pi}K_{\mu}K^{\mu}+\frac{1}{32\, \tilde{e}^2}\text{Tr}\left(\left[ K_{\mu}, K_{\nu}\right] \left[ K^{\mu}, K^{\nu}\right]\right) \, , 
\label{eq:Lagden}
\end{equation}
where we have introduced the $SU(2)$ valued Lorentz four-vector ${K_{\mu}\,=\,U^{-1}\partial_{\mu} U}$, 
and $U$ is an $SU(2)$ valued Lorentz scalar referred to as the Skyrme field. In the following sections we will work with $\tilde{e}=8\pi G e$ where $e$ is a dimensionless coupling.

\subsection{Merons}
\label{sec:merons}
As a result of the work of the authors in~\cite{Canfora:2013osa,Ayon-Beato:2015eca,Canfora:2017yio,Canfora:2018ppu,Ayon-Beato:2019tvu} the authors of~\cite{Ipinza:2020xgc} have shown that the Einstein-Skyrme theory~\eqref{eq:skyrmeAction} has the same equations of motion as a particular Einstein-Yang-Mills (EYM) theory. The latter contains a massive non-Abelian gauge field with mass $m$ in a pure gauge configuration $A_\mu = \lambda U^{-1}\partial_\mu U$, with the $SU(2)$ group element $U$, and the real-valued parameter $\lambda\neq 0,\,1$. Such configurations have been known for over 30 years as solutions to classical $SU(2)$ Yang-Mills equations and are referred to as {\it merons}~\cite{RevModPhys.51.461}. Merons are classical topological soliton solutions. 

The action of this theory is given by
\begin{equation}\label{eq:meronAction}
 S_{\text{meron}} = \int d^5x \sqrt{-g}
 \left (
 \frac{1}{16\pi G} (R-2\Lambda) 
 + \frac{1}{16\pi \gamma^2} \text{Tr}[
 F_{\mu\nu} F^{\mu\nu} - 2 m^2 A^\mu A_\nu
 ]
 \right) \, ,
\end{equation}
with the Yang-Mills coupling constant $\gamma$, the cosmological constant $\Lambda=-6/L^2$ with $L$ the $AdS$ radius and the Proca mass $m$~\cite{Kunimasa:1967GT,Shizuya:1975ek}. 
In this paper we consider massless gauge fields, $m=0$, except in section~\ref{sec:staticBBs}, where the massive solution serves as a consistency check as discussed there. 

For particular values of $\lambda$, the meron and the Skyrmion theory have the same solutions under the identifications 
\begin{equation}
    A_\mu=\lambda K_\mu \, , \quad 
    F_{\mu\nu} = \lambda (\lambda-1) [K_\mu,K_\nu] \, , \quad 
    \frac{m^2 \lambda^2}{\pi \gamma^2} = \frac{f_\pi^2}{2}\, , \quad 
    \frac{\lambda^2(\lambda-1)^2}{\pi \gamma^2} = 
    \frac{1}{2 \tilde{e}^2}\, ,\label{eq:identify}
\end{equation}
where $f_\pi$ is the coupling constant multiplying the kinetic Skyrmion term 
in the action eq.~\eqref{eq:skyrmeAction}.

In order for the meron solution to carry the appropriate half-integer topological charge, we will see in the next subsection that it is crucial that $|\lambda|=1/2$. Originally, merons were considered in Euclidean space-time and interpreted as {\it half-instantons}.\footnote{Instantons 
carry integer topological charge, merons carry half-integer topological charge.} 
By definition $U^{-1}\partial U$ is a pure gauge solution. Thus, merons with $A=\frac{1}{2} K =\frac{1}{2} U^{-1}\partial U$ are {\it half of pure gauge solutions}. 
Note that instantons are pure gauge only on the compact surface at infinity, while merons are half of pure gauge solutions everywhere. Also, instantons have a size while merons do not~\cite{RevModPhys.51.461}.

\subsection{Topological charges}
\label{sec:topologicalCharges}
Our goal is to obtain a holographic field theory interpretation of topological solutions in four dimensions via a five dimensional gravitational theory with Lorentzian signature. 
In five dimensions topological invariants such as the second Chern number may be defined on any four-dimensional submanifold~\cite{Manton:2004tk}. Hence to begin, we now consider pure gauge solutions to Yang-Mills theory in flat Euclidean space-time, i.e. in the Euclideanized 3+1-dimensional field theory living on the boundary of $AdS_5$. The spatial part $\mathbb{R}^{3}$ is topologically equivalent to $S^3$ given certain conditions which are fulfilled in our case.\footnote{More precisely, the stereographic projection of $\mathbb{R}^{n}$ with a point added at infinity is an $\mathbb{R}^n$ representation of $S^n$. Consider the definition of a based map, such as our gauge field solution, $A$: It maps one manifold $\mathcal{M}$ to another manifold $\bar{\mathcal{M}}$, 
$A: \mathcal{M}\to \bar{\mathcal{M}}$, and identifies a base point $x_0\in\mathcal{M}$ with a point $y_0\in \bar{\mathcal{M}}$ by $A(x_0)=y_0$. 
It is equivalent to another based map, $A_0: \mathbb{R}^n\to \bar{\mathcal{M}}$, if 
$\lim\limits_{x \to \infty} A(x)\to y_0$. 
Choosing the point at infinity as our base point, the maps $A$ and $A_0$ are equivalent. This shows that our gauge field can be considered as a based map $A:\, S^3\to S^3$.
}
In our case, the gauge field solutions mapping $A:\mathbb{R}^3 \to S^3$ are equivalent to those mapping $A:S^3 \to S^3$. 

In the previous subsection, we have learned that Skyrmion solutions are equivalent to meron solutions. Both Skyrmions and merons carry topological charge~\cite{RevModPhys.51.461,Manton:2004tk}. In order to reveal their topological nature, we take the spatial $\mathbb{R}^3$ part of a given space-time and compactify it to $S^3$ by adding a point at spatial infinity. Then the solutions map a point on that coordinate space three-sphere, $S^3$, to a point on another three-sphere, $S^3_{SU(2)}$, which is the group manifold of the Lie group $SU(2)$. In other words, the  solutions are maps of $S^3$ onto itself. All such maps fall into homotopy classes labeled by a topological charge $\mathpzc{q}$. 
Topologically distinct solutions ``wrap'' the $S^3_{SU(2)}$ around the $S^3$ a different number of $\mathpzc{q}$ times. In this subsection, we consider the topological properties of Skyrmions and merons in the context of an $SU(2)$ Yang-Mills theory.

Skyrmions and merons are topological soliton solutions in field theories~\cite{RevModPhys.51.461,Manton:2004tk}. 
It has also been known for a while, that solitons in $D$ dimensions can be interpreted as instantons in $D+1$ dimensions.

\paragraph{A simple toy model:} 
In order to get an intuition for these topological relations, it may help to envision the analogous $U(1)$ problem~\cite{RevModPhys.51.461,Manton:2004tk}. In that case, each previously considered $S^3$ becomes an $S^1$, also known as a circle. The topological solution maps the spatial circle parametrized by an angle $\psi$ to the gauge manifold circle, parametrized by an angle $\theta$, also known as the phase angle. Then the topological charge would be given by 
\begin{equation}
\mathpzc{q}_{circle} =
{\frac{1}{2\pi}\oint d\Omega_i J^i}
=-\frac{i}{2\pi}\oint d\Omega_i \epsilon^{ij}(\partial_j \omega) \omega^{-1} = \frac{n}{2\pi} \oint d\Omega_i \epsilon^{ij} \partial_j \theta \, ,    
\end{equation} 
with the topological current, $J^i=-i \epsilon^{ij} (\partial_j \omega) \, \omega^{-1}$, for a pure gauge field, $\omega=e^{in\theta}$, $\theta\in [0,2\pi]$, $x_i = (x,y) = r(\text{cos}\psi,\text{sin}\psi)$, $\psi\in [0,2\pi]$, and $d\Omega_i=\hat{n}_i d\psi$ with the outward normal vector $\hat{n}_i =(\text{cos}\psi,\text{sin}\psi)$, on the $S^1$. Then, 
\begin{equation}
\mathpzc{q}_{circle} = \frac{1}{2\pi}n \int\limits_0^{2\pi} d\theta = n \, .
\end{equation}
The integer $n\in \mathbb{Z}$
is the winding number of the $S^1_{U(1)}$, which parametrizes the $U(1)$ gauge manifold, around the spatial circle, $S^1$. %

\paragraph{Topological charges of $SU(2)$ Yang-Mills solutions in four dimensions:} 
In a four-dimensional Euclidean $SU(2)$ theory with Lagrangian $\mathcal{L}=\frac{1}{4} F_{ij}^a F^{ija}$ where $F=dA+A\wedge A$ with Yang-Mills gauge coupling $g_{\text{\tiny YM}}$, a topological charge of the Euclidean field configuration (the solution to the equations of motion) can be defined by\footnote{Generally, this equation would include a limit instructing us to evaluate the expression on the boundary at infinity, and gauge field configurations would  have to be pure gauge only in that limit. Since we consider a more restrictive case, pure gauge solutions everywhere in space-time, we drop this limit in all of the expressions for the topological charge.}
\begin{equation}\label{eq:instantonCharge}
    \mathpzc{q}
    = \frac{{g_{\text{\tiny YM}}}^2}{8\pi^2} \int D d^4 x = 
    {\frac{1}{16\pi^2}}
    \oint\limits_{S^3} \hat n_i J^i d^3 x \, ,
\end{equation}
where $\hat n$ is the timelike outward normal vector on a compact Cauchy surface, the spatial coordinate space $S^3$ mentioned above. The topological current 
\begin{equation}\label{eq:topoCurrent}
    J_i = {g_{\text{\tiny YM}}}^2 \epsilon_i{}^{jmn} \left [ 
    A_j^a \partial_m A_n^a + \frac{g_{\text{\tiny YM}}}{3} \epsilon_{abc} A_j^a A_m^b A_n^c
    \right ] \, ,
\end{equation}
has a divergence which is a pseudoscalar density defined as
\begin{equation}
    D=\frac{1}{4} F_{ij}^a \tilde F_{ij}^a = \frac{1}{2{g_{\text{\tiny YM}}}^2} \partial_i J^i\, .
\end{equation}
Later we will see that our $AdS_5$ solutions induce pure gauge solutions in the four-dimensional dual field theory. 
It can be shown that pure gauge fields in four dimensions carry topological charge $\mathpzc{q}\in \mathbb{Z}$ 
given that their energies decrease fast enough at infinity~\cite{Manton:2004tk}.
This can be seen by writing eq.~\eqref{eq:topoCurrent} in pure gauge considering that $A_i \to A_i' = \omega A_i \omega^{-1} - (i/g_{\text{\tiny YM}})(\partial_i \omega)\omega^{-1}$, and noting that under such a gauge transformation the current transforms as\footnote{The trace over generators is related in the usual way to the structure constant of $SU(2)$, $\text{tr}(\tau_a \tau_b\tau_c)=\frac{i}{2} \epsilon_{abc}$.}
\begin{equation}\label{eq:topoCurrentTrafo}
    J_i \to J_i' = J_i + 2 i g_{\text{\tiny YM}} \epsilon_i{}^{jmn} \text{tr}\{ \partial_j(\partial_m \omega A_n \omega^{-1}) \}-\frac{2}{3} \epsilon_i{}^{jmn} \text{tr}\{ \partial_j \omega \omega^{-1} \partial_m \omega \omega^{-1} \partial_n \omega \omega^{-1} \} \, .
\end{equation}
Taking only the pure gauge part of~\eqref{eq:topoCurrentTrafo}, we get the current of a pure gauge field $A=(\partial \omega) \omega^{-1}$.  
Hence the winding number $\mathpzc{q}$ is a property of the pure gauge field $A$.  
For the $SU(2)$ Yang-Mills theory the pure gauge part of eq.~\eqref{eq:topoCurrentTrafo} happens to be the definition of the degree of a map from $S^3$ to $SU(2)$ and hence takes on integer values~\cite{Manton:2004tk}. 
In this case, the topological charge~\eqref{eq:instantonCharge} is $\mathpzc{q}=c_2 =N$, where the second Chern number of the gauge field solution is $c_2 = \int C_2$ (with the second Chern form $C_2=\frac{1}{8\pi^2} \left(\text{tr}(F\wedge F)-\text{tr}F\wedge \text{tr}F\right)$), 
and the instanton number is $N=-\frac{1}{8\pi^2}\int \text{tr} (F_{ij} \star F_{ij}) d^4x$~\cite{Manton:2004tk}. 

The meron topological charge identifying it as a half-instanton~\cite{RevModPhys.51.461} is also defined by eq.~\eqref{eq:instantonCharge}. Evaluating the gauge transformation of the current~\eqref{eq:topoCurrent} for a half of a pure gauge field $1/2 (\partial \omega) \omega^{-1}$, we get one half of the pure gauge current of the pure gauge field $A$ yielding~\cite{RevModPhys.51.461}
\begin{equation}\label{eq:topoCurrentTrafoMeron}
    J_i\left(\frac{A}{2}\right) \to J_i'\left(\frac{A'}{2}\right) =\frac{1}{2}\times J_i'(A')
    {=}\frac{1}{2}\times\frac{-2}{3} \epsilon_i{}^{jmn} \text{tr}\left [ \partial_j \omega \omega^{-1} \partial_m \omega \omega^{-1} \partial_n \omega \omega^{-1} \right ]\, .
\end{equation}
Using the current~\eqref{eq:topoCurrentTrafoMeron} in eq.~\eqref{eq:instantonCharge} implies that the meron, $A_i$, as a half of a pure gauge solution has half the topological charge of the corresponding pure gauge solution. 

As a topological soliton a Skyrmion in $D$-dimensions carries a topological charge referred to as the {\it baryon number} for historical reasons~\cite{Manton:2004tk,Ipinza:2020xgc}. This baryon number is exactly the topological charge $\mathpzc{q}$. 
Considering the Skyrmion and the meron solutions as topological solitons in four dimensions and considering their topological charge~\eqref{eq:instantonCharge}, we find $\mathpzc{q}_\text{meron}=\frac{1}{2}\times \mathpzc{q}_\text{\,Skyrmion}$. 

While the discussion of this section has been mainly concerned with gauge field configurations, the inclusion of these configurations as gravitating matter contributions has also been considered. Fully backreacted analytic solutions within the Einstein-Skyrme model with topological charge $\pm 1$ were originally found in~\cite{Ayon-Beato:2015eca}. These solutions were further investigated/extended in~\cite{Canfora:2017yio} where the authors find that Yang-Mills configurations which would have been singular in flat space are ``regularized'' by their coupling to the surrounding geometry. The authors name this effect the gravitational catalysis of merons~\cite{Canfora:2017yio}. Solutions with higher baryon number have also been considered in the context of four dimensional $SU(3)$-Skyrme-Einstein theory with cosmological constant~\cite{Ayon-Beato:2019tvu}. There the authors presented the first analytic self-gravitating Skyrmions with baryon charge 4 in four dimensions and find a novel transition at nonzero Baryon charge between embedded and non-embedded gauge field configurations. The topological solution presented in our work bears a resemblance to those constructed in~\cite{Canfora:2018ppu}. There the authors work with an Einstein-Yang-Mill's model with non-zero cosmological constant and construct a black hole solution with topological charge 1 and discuss its thermodynamics. The solution we find is distinguished by the admission of arbitrary $q\in \mathbb{N}$. Furthermore we also present two additional analytic black brane solutions with zero topological charge; a  Reissner-Nordstr\"om like solution and a rotating solution.

\section{Skyrmion black brane/hole solutions \& thermodynamics}
\label{sec:blackBraneSolutions}
In this section we show that analytic solutions of the field equations can be obtained for the metric tensor elements which describe static or rotating black branes/holes with hair in a five-dimensional universe which is asymptotically $AdS_5$.  For the static black branes/holes analytic solutions can be obtained for scalar fields (hair) which are either massless or massive $SU(2)$-valued Skyrmion fields. For the case of a static black brane with massless scalar hair, the solution uniquely specifies an unknown function $\chi(r)$, which determines the Skyrmion field.  For the rotating black brane the Skyrme field is chosen to be a standard, hedge-hog form~\cite{Manton:2004tk}. A third class of solutions, asymptotic to global $AdS_5$, is found to be topologically non-trivial.
The thermodynamic properties of these solutions are computed and discussed.

\subsection{Static Skyrme-$AdS_5$ black branes}
\label{sec:staticBBs}
First, we find a solution which extremizes the Skyrmion action~\eqref{eq:skyrmeAction} with vanishing pion coupling, $f_\pi=0$.  
Below, we will see that the static solution we are about to present carries an oscillating charge driven by an oscillating chemical potential and takes the same form as an asymptotically $AdS_5$ Reissner-Nordstr\"om black brane. Hence, we choose the informed ansatz for the invariant line element accordingly 
\be 
ds^2=-A(r) dt^2+  \frac{dr^2}{A(r)}+ \frac{r^2}{L^2} \exd x^i\exd x_i\, .  \label{eq:Mansatz}
\ee
In this expression $r$ is the radial coordinate with the $AdS$-boundary at $r=\infty$, $t$ is the temporal field theory direction and $x_i=(x_1,x_2,x_3)$ are the remaining field theory directions. 
A field rotating around the $z-$direction in internal space with angular velocity $\omega$ is described by 
\be
U \,=\, n_1(r)  \mathds{1} + i n_2(r) \left( \tau_1 \cos( - \omega t) + \tau_2 \sin( -\omega t) \right) + i n_3(r) \tau_3 \, , \label{eq:Sansatz}
\ee
where $\mathds{1} $ is the $2\times 2$ unit matrix, $\tau_{1,2,3}$ are the Pauli matrices $\tau_1=((0,1),(1,0)),\, \tau_2=((0,-i),(i,0))$ and $\tau_3=((1,0),(0,-1))$, such that $\text{tr}\{\tau_1 \tau_2 \tau_3\} = 2 i$ 
, and  $\vec n=(n_1,n_2,n_3)$ is a unit vector whose components satisfy the condition
\be 
\sum_i\,n_i^2\,=\,1\, .
\ee
The $n_i$ are restricted to being functions of only the radial coordinate $r$.  A simple choice for the $n_i$'s is,
\be \label{eq:n1n2Ansatz}
n_1(r)\,=\,\cos(\chi(r))\, ,\hspace{.5cm} n_2(r)\,=\, \sin(\chi(r))\, ,\hspace{.5cm}  n_3(r)=0.
\ee
This is one of the choices for $n_1, n_2$ and $n_3$ analyzed in  \cite{Ioannidou:2006nn}.

The Einstein equations reduce to two independent equations while the variation of the action with respect to $\chi(r)$ yields a single equation, 
\begin{align}
    -\frac{3 A'(r)}{2r}-\frac{3 A(r )}{r^2}+\frac{\omega ^2}{2e^2} \chi '(r)^2 \sin ^2(\chi (r))+6&=0, \label{eqn:skyrme}\\
    r^2 A''(r)+\frac{r^2\omega ^2}{e^2} \chi '(r)^2 \sin ^2(\chi (r))-12r^2+4r A'(r)+2A(r)&=0, \label{eqn:A} \\
    3 \sin(\chi(r))\chi'(r)+r \cos(\chi(r))\chi'(r)^2+r \sin(\chi(r))\chi''(r)&=0. \label{eq:skeom}
\end{align}
We can solve eq. (\ref{eq:skeom}) separately from the metric sector and find,
\begin{equation}
  \chi(r)=  \arccos\left(-\frac{\chi_2}{2r^2}-\chi_1\right). \label{eq:SkyrmionSol}
\end{equation}
Inserting the solution into eq.\ (\ref{eqn:skyrme}) and eq.\ (\ref{eqn:A}) we find,
\begin{equation}\label{eq:SkyrmionSolA}
   A(r)= -\frac{a_1}{r^2}-\frac{ \chi_2^2 \omega ^2}{6e^2 r^4}+r^2. 
\end{equation}
The solution for the blackening factor is strongly reminiscent of an asymptotically $AdS_5$ Reissner-Nordstr\"om black brane. 
However, the solution found here is distinguished as it has only one horizon and hence there exists no extremal solution.
Taking the limit of $\chi_2\rightarrow 0$ takes us back to an $AdS_5$ Schwarzschild black brane, and hence we identify 
$a_1=m_s$ 
as the mass of the black brane. We therefore take the solution for the blackening factor to be,
\begin{equation}
      A(r)= -\frac{
      m_s
      }{r^2}-\frac{ \chi_2^2 \omega ^2}{6e^2 r^4}+\frac{r^2}{L^2}.  \label{eq:blackening_factor}
\end{equation}
Interestingly we find that the term proportional to $r^{-4}$ typically associated with the charge of the RN black brane depends on $\omega$, the angular velocity of rotation in the internal $SU(2)$ space. 

As discussed above, these Skyrmion solutions can be written as meron solutions, i.e.~half pure gauge solutions. We have explicitly checked this claim  for the solution given here. Under the identification $A_\mu = - 1/2 \,  U^{-1}\partial_\mu U$, the Einstein-Yang-Mills equations of motion are satisfied. 
As expected from the identifications~\eqref{eq:identify}, this is a solution for massless gauge fields, i.e.~$m=0$. 

Now we confirm that our choice of a vanishing kinetic term was not pathological. Suppose we had included the kinetic term by choosing $f_\pi\neq 0$ in the Skyrme action~\eqref{eq:skyrmeAction} 
, which implies a non-vanishing Proca mass for the meron, $m\neq 0$.  
Then we would have found that Einstein's equations reduce to three independent equations, and the Skyrme field equation provides a fourth independent equation.
Solving all four equations simultaneously leads to two possible solutions, one of these solutions is smoothly connected to the solution displayed in eq.\ (\ref{eq:blackening_factor}) and eq.\ (\ref{eq:SkyrmionSol}) and is given by,
\begin{equation}\label{eq:blackening_factor_kinetic}
    A(r)=\frac{r^2}{L^2}-\frac{m_s}{r^2},\quad \chi(r)=k \pi,\quad \forall k\in\mathbb{Z} \, .
\end{equation}
This can be seen by taking the limit of $\chi_2\rightarrow 0$ while taking $\chi_1=(-1)^{k+1}$.
As with our non-trivial solution eq.\ (\ref{eq:blackening_factor}) without a kinetic term we have explicitly checked that 
the solution eq.\ (\ref{eq:blackening_factor_kinetic}) can be written as the solution for a massive meron. This concludes our analysis of the massive meron case and we will consider only massless merons, $m=0$, from here on.

\begin{figure}[htb]
    \begin{subfigure}[b]{0.5\textwidth}
    \includegraphics[width=80mm,scale=0.5]{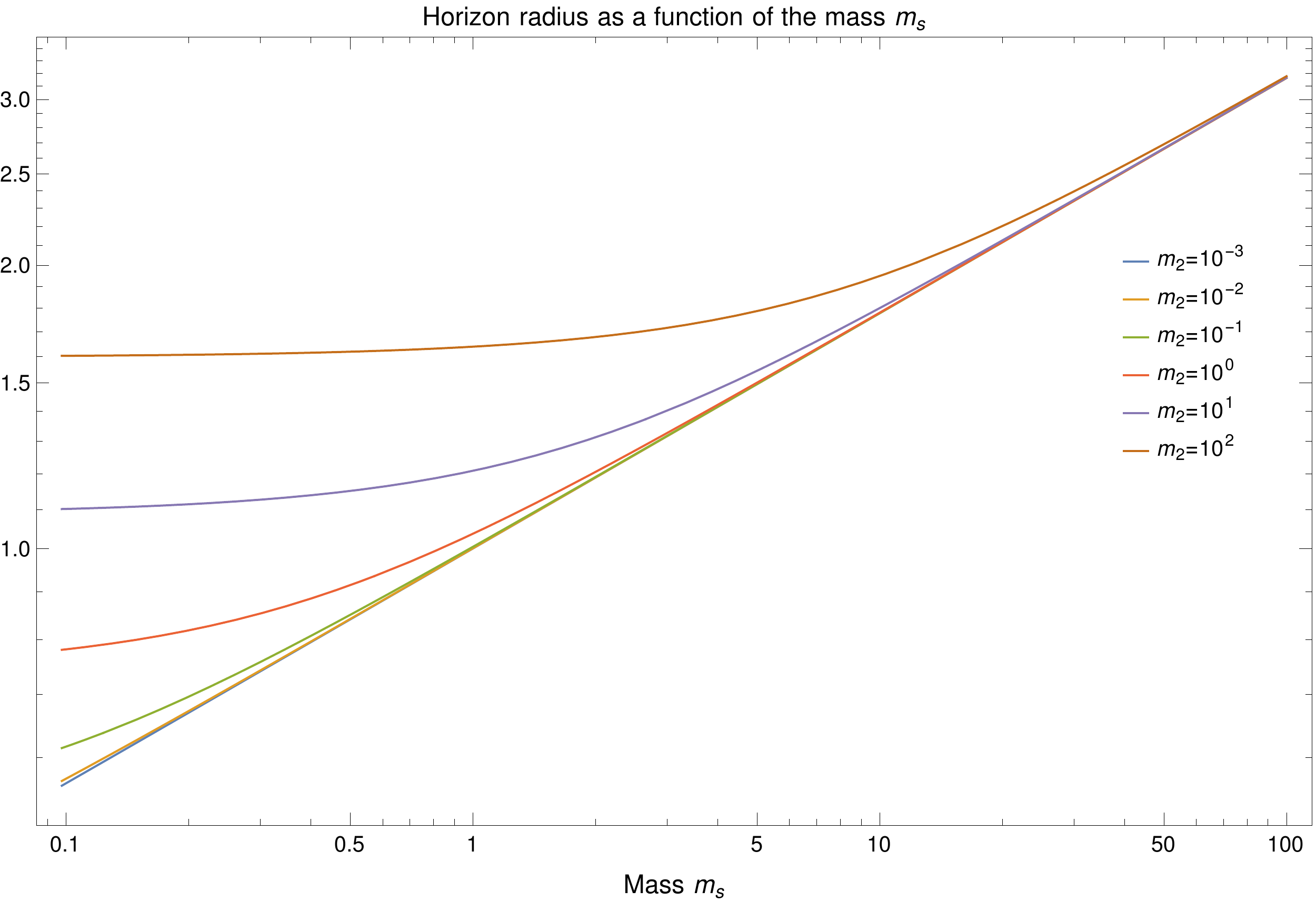}
    \end{subfigure}
    \begin{subfigure}[b]{0.5\textwidth}
    \includegraphics[width=80mm,scale=0.5]{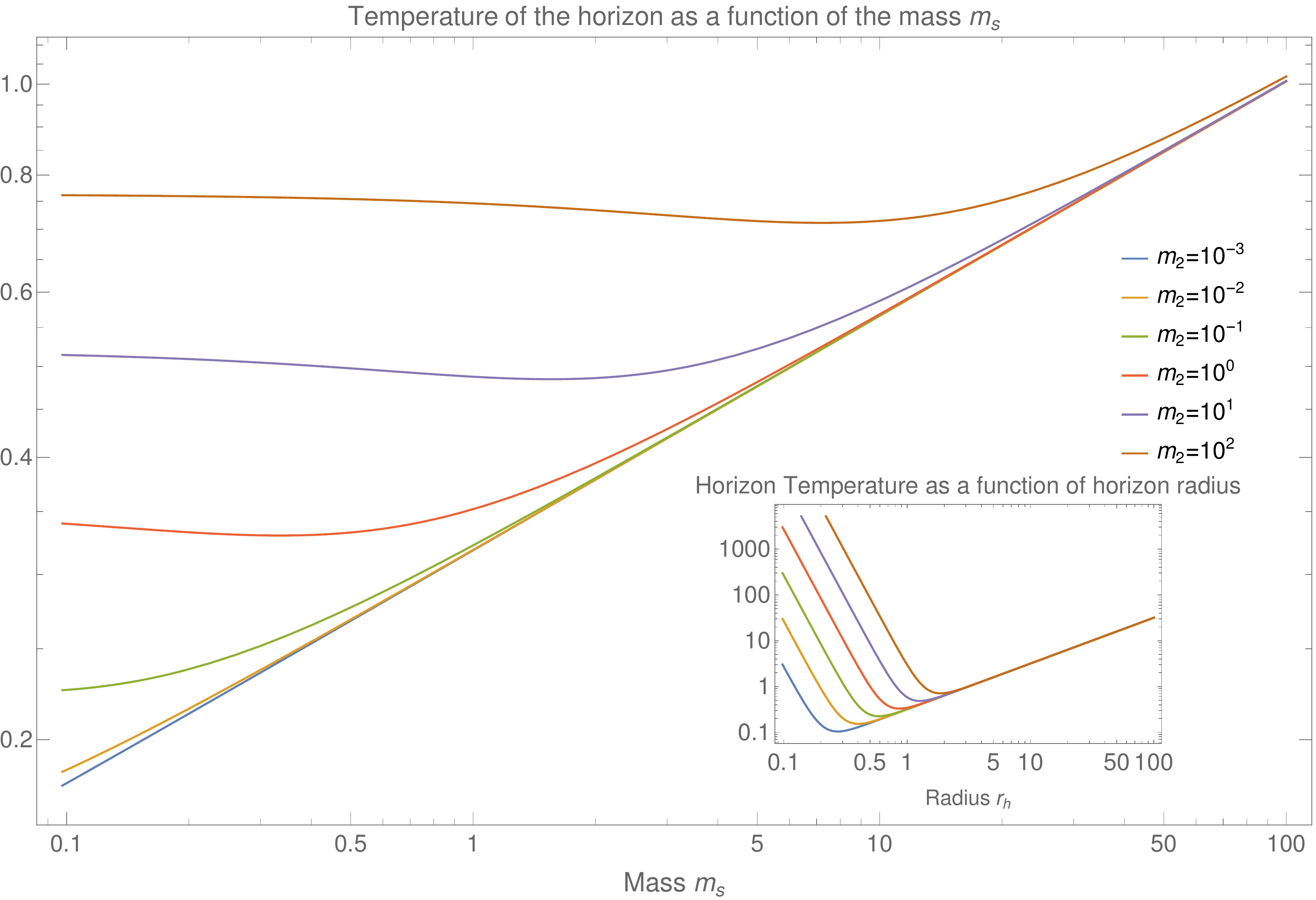}
    \end{subfigure}
    \caption{\textit{Left:} The horizon radius, $r_h$ as a function of the mass $m_s$, is displayed on a log-log plot. \textit{Right:} The temperature $T$ as a function of the mass $m_s$ is displayed on a log-log plot. The inset displays the temperature $T$ as function of the horizon radius $r_h$. All images are displayed with the additional Skyrmion mass parameter $m_2=(\omega\chi_2/e)^2$ scanned over a window spanning five orders of magnitude with $m_2=(10^{-3},10^{-2},10^{-1},10^{0},10^{1},10^{2})$ and the $AdS$ radius $L=1$. Between the left and right images the colors correspond to the same value of the additional Skyrmion mass parameter $m_2$. 
    \label{fig:temperature}}
\end{figure}

Returning to our $m=0$ solution, eq.~\eqref{eq:blackening_factor}, we begin by addressing the location of the horizon. This can be obtained via the location of the poles of $G_{rr}$, the radial component of the metric. 
As already mentioned the solution is reminiscent of a  Reissner-Nordstr\"om black brane although there is only a single horizon. There are six solutions to $A(r_h)=0$; we take the root with $Im(r_h)=0$ and $r_h>0$,
\begin{equation}
r_h= \frac{1}{6^{1/3}}\left(\frac{ \left(\sqrt{9 L^4 m_2^2-48 L^6 m_s^3}+3 L^2 m_2\right)^{2/3}+ (48 L^6 m_s^3)^{1/3}}{(\sqrt{9 L^4 m_2^2-48 L^6 m_s^3}+3 L^2 m_2)^{1/3}               }\right)^{1/2}.    \label{eq:RN_rh}
\end{equation}
Where we have put $m_2=(\chi_2\omega/e)^2$. Taking the limit of vanishing mass $m_s\rightarrow 0$ does not lead to an empty $AdS$ solution, it instead leads to an additional black brane blackening factor and horizon radius of,
\begin{equation}
    A(r)=\frac{r^2}{L^2}-\frac{m_2}{6r^4} \, ,\qquad r'_h=\left(\frac{m_2L^2 }{ 6}\right)^{1/6} \, ,
\end{equation}
where we note that the product $(\chi_2\omega/e)^2$ now acts like a mass, revealing the origin of our choice of labeling.
We can calculate the temperature of both solutions via standard methods,
\begin{equation}
    T=\frac{1}{4\pi}\left|A'(r_h)\right|  \, . \label{eq:RNtemp}
\end{equation}
Although we do not display the explicit formula for the temperature, in figure~\ref{fig:temperature} we display eq.\ (\ref{eq:RNtemp}) for various values of $\omega$ and $\chi_2$ for fixed $m_s$. We can see the temperature grows non-linearly for small $\chi_2$ but approaches linear behavior for large $\chi$. 

The horizon area can be computed and associated to the entropy via $S=A/4G_5$. Computing the area $A$, we find,
\begin{equation}
    A=\int \sqrt{\text{det}(g_{ij}(r_h))}=\frac{r_h^3}{L^3}\int \exd^3 x=\frac{r_h^3}{L^3} V,
\end{equation}
where $V=\int \exd^3x$ is an infinite Euclidean volume. We then find the entropy density to be given by,
\begin{equation}\label{eq:sStatic}
    s=S/V=\frac{1}{4G_5}\frac{r_h^3}{L^3}\, ,
\end{equation}
with $r_h$ given by eq.\ (\ref{eq:RN_rh}). The entropy density scales with horizon radius in the same way that both the Schwarzschild and Reissner-Nordstr\"om black branes scale. 

\begin{figure}
    \centering
    \includegraphics[width=14cm]{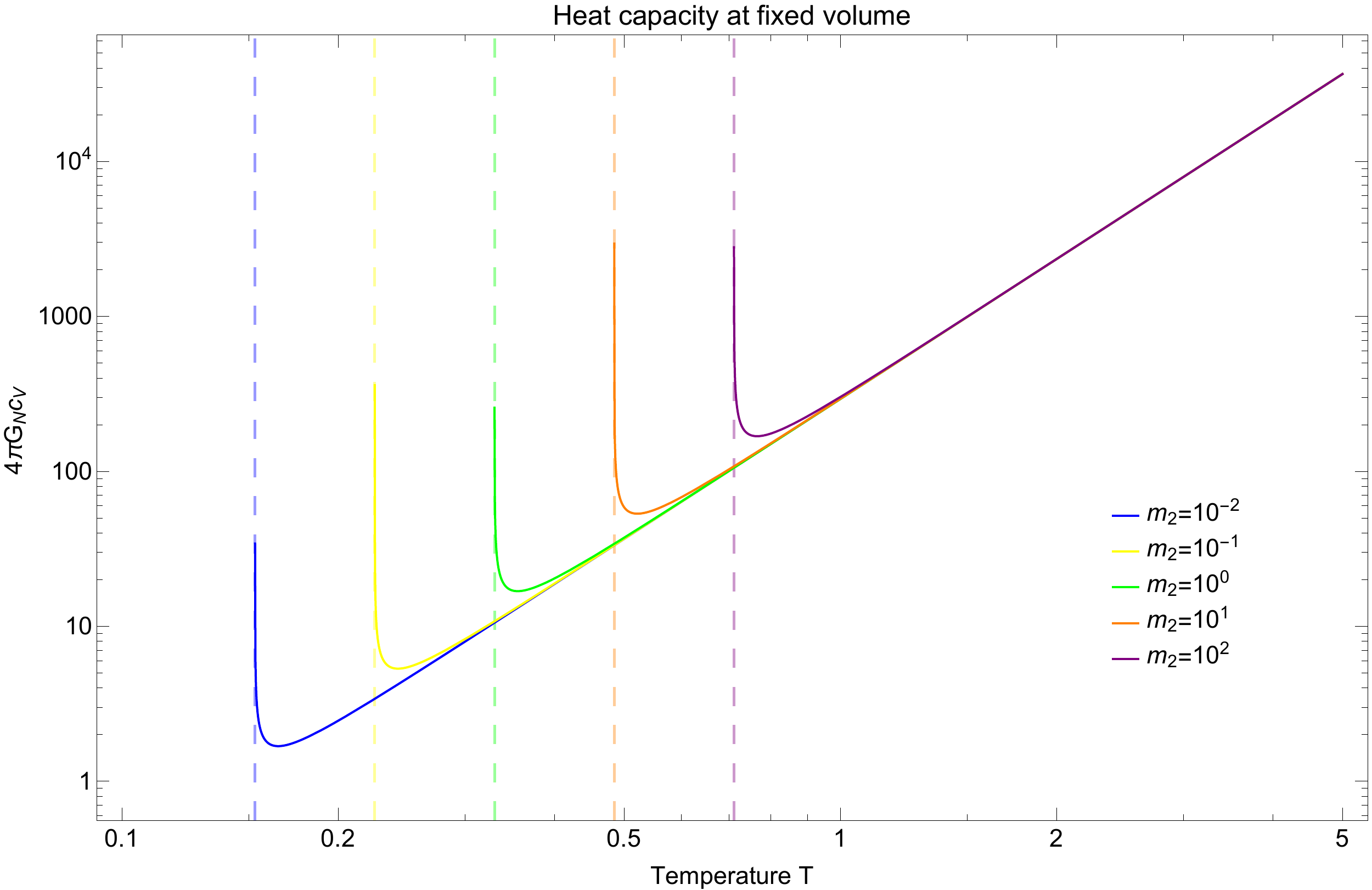}
    \caption{
    \textit{ Heat capacity of the static Skyrme-$AdS_5$ black branes.} Displayed here is the heat capacity at fixed volume of the Reissner-Nordstr\"om-like solution, with the blackening factor given by eq.~\eqref{eq:blackening_factor}, described in section~\ref{sec:blackBraneSolutions}. We display this quantity over several orders of magnitude of the Skyrme mass parameter $m_2=(10^{-2},10^{-1},10^0,10^1,10^2)$. The vertical lines displayed indicate the lower bound on the temperature as given in eq.\ (\ref{eq:T_bound}). We recall that the mass parameter is given by $m_2=(\chi_2\omega/e)^2$.
    \label{fig:HeatCap_RN}}
\end{figure}

We consider the thermodynamic stability of our solutions following the standard techniques. 
Requiring the second variation of the entropy with respect to the temperature to be negative, or equivalently requiring the heat capacity to be positive, implies stability of the system against thermal flucuations~\cite{Chamblin:1999hg},
\begin{equation}
   c_V=\left( \frac{\partial E}{\partial T}\right)_V\geq 0\, .
\end{equation} 
Likewise requiring the second variation of the entropy with respect to the charge density to be negative, or equivalently requiring the charge susceptiblity to be positive, implies thermodynamic stability against charge fluctuations,
\begin{equation}
    \chi_{\rho} =\left( \frac{\partial \mu}{\partial \rho}\right)_V\geq 0\, .
\end{equation}
Using the chain rule and holding the volume $V$ fixed in all derivatives, we may conveniently write
\begin{align}
    c_V&=\left ( \frac{\partial E}{\partial T}\right )_V=\frac{\partial M}{\partial m_s}\frac{\partial m_s}{\partial r_h}\frac{\partial r_h}{\partial T}=\left(\frac{3}{4\bar{\kappa}}\right)\left(\frac{m_2}{3 r_h^3}+4 \frac{r_h^3}{L^2}\right)\left(\frac{1}{\pi  L^2}-\frac{5 m_2}{12 \pi  r_h^6}\right)^{-1} \\
    &=\frac{1}{\bar{\kappa}}\frac{3 \pi  r_h^3 \left(L^2m_2+12 r_h^6\right)}{12 r_h^6-5 m_2L^2}  \, ,
    \end{align}
for the static Skyrmion black brane solution. It is evident from this expression that the heat capacity would only be negative in two cases. The first case is $m_2<0$. This is ruled out because $m_2$ is defined as a square $m_2=(\chi_2\omega/e)^2$ and hence is positive or zero. 
The second case, namely that $m_2>12 r_h^6/5L^2$, provides a bound on $m_2$. In fact the saturation of this bound, $m_2=12 r_h^6/5L^2$ can be reduced to,
\begin{equation}
   m_2= 4 \sqrt{\frac{5}{3}} L m_s^{3/2}\, .\label{eq:Range}
\end{equation}
At the saturation point the heat capacity becomes infinite and the horizon radius is given, in terms of $m$, as,
\begin{equation}
  r_h=  \frac{\sqrt{(\sqrt{5}+2)^{1/3}+\sqrt{5}+2} \sqrt{L} m_s^{1/4}}{3^{1/4} (\sqrt{5}+2)^{1/3}}\, .
\end{equation}
The bound on the horizon radius provides a lower bound on the temperature given, in terms of $m_2$, as,
\begin{equation}
   T_{min}= \frac{2^{2/3} 3^{5/6} m_2^{1/6}}{5^{5/6} \pi } \, .\label{eq:T_bound}
\end{equation}
It is interesting to note that this system has no extremal limit in which $T\rightarrow 0$. This curious behavior can be traced back to the sign of the term containing $m_2$ (within $\chi_2$) in the blackening factor in eq.\ (\ref{eq:blackening_factor}). Instead of approaching zero temperature for a unique value of the parameters $m_s$ and $m_2$, the sign of the term including $m_2$ leads to a minimum temperature solution. In addition the heat capacity reveals that an infinite amount of energy must be supplied to reach this minimum temperature.  Figure~\ref{fig:HeatCap_RN} displays the heat capacity for various values of $m_2=10^{-2}, ..., 10^2$. We can see that the curves truncate at the bound given by eq.\ (\ref{eq:Range}) displayed as vertical lines at the location of $T_{min}$. 

The charge susceptibility $\chi_\rho$ is a tensor in isospin space, $ {(\chi_{\rho})}^a_b$, defined as,
\begin{equation}
    {(\chi_{\rho})}^a_b=\frac{\partial Q^a}{\partial \mu^b}\, ,
\end{equation}
where $\rho$ here is a label and $a,b=1,2,3$ are the directions in isospin space. This then defines a relation between the charge density and chemical potential  $Q^a=\vev{J^a}= {(\chi_{\rho})}^a_b\mu^b$. 
Expressing the isospin charge in this way we find,
\begin{equation}\label{eq:Static_sus}
 {(\chi_{\rho})}^a_b=\begin{pmatrix}
 \frac{\left(1-2 \chi _1^2\right) \chi _2}{2 \chi _1-2 \chi _1^3} & 0 & 0 \\
 0 & \frac{\left(1-2 \chi _1^2\right) \chi _2}{2 \chi _1-2 \chi _1^3} & 0 \\
 0 & 0 & \frac{\chi _1 \chi _2}{\chi _1^2-1} \\
\end{pmatrix}\, .
\end{equation}
Analyzing the entries of the susceptibility tensor we find that,
\begin{equation}
   (\chi_\rho)^a_b>0 \quad \text{if}\quad
  \begin{cases}
        1>\chi_1>1/\sqrt{2} & \text{\&} \quad\chi_2<0 \\
        -1<\chi_1<-1/\sqrt{2} &\text{ \&} \quad \chi_2>0
    \end{cases}
\end{equation}
Considering our theory with these parameter restrictions ensures that our solution is stable against isospin charge fluctuations.

\subsection{Rotating Skyrme-$AdS_5$ black branes}
\label{sec:rotatingBBs}
Solutions of Einstein's equations describing black holes with scalar hair have been found in three and four dimensions  \cite{Henneaux:2002wm,Banados:2005hm,Brihaye:2013tra,Anabalon:2016izw}.
In  \cite{Harms:2016pow,Harms:2017yko}, 
new types of such asymptotically $AdS$ solutions  in three space-time dimensions were found for rotating black holes.  Their matter contribution is the non-linear  $\sigma$-model, and they describe both self-gravitating topological solitons and  black holes with hair.  One of the results of the analysis of these models was the discovery of a hidden symmetry which allowed some of the metric tensor elements to be expressed in analytic forms, and which lead to a family of black hole solutions . In \cite{Harms:2019cag} this symmetry was shown to exist in a three-dimensional space-time with a dilatonic black hole and  in four-dimensional models with and without matter.  In all of these cases the imposition of the constraint 
\be 
g_{t t}(r,p)\,=\,\omega^2 g_{\phi \phi}(r,p)+ 2\,\omega g_{t \phi}(r,p) \, ,
\label{eq:constraint}
\ee
where $p\,=\,a\,\cos(\theta)$, 
led to a reduction in the number of independent field equations and to analytic solutions for some of the metric tensor elements.  The number of independent field equations after the imposition of the constraint is one less than the number of independent fields, allowing the freedom to specify one of the undetermined metric tensor elements, i.e. one not obtained in analytic form, or to specify the matter field. 
In this section we show that analytic solutions for the metric tensor elements of a black brane in a five-dimensional space-time which is asymptotically $AdS$ can also be obtained for rotating branes. In order to obtain analytic solutions we make use of the constraint given in eq.\ (\ref{eq:constraint}). We comment further on the significance of the constraint~\eqref{eq:constraint} and the reduction of field equations in the discussion section~\ref{sec:Discussion}.

The Lagrangian density is again chosen to be of the form in eq.(\ref{eq:Lagden}) 
with vanishing pion coupling, $f_\pi=0$,  
and the field $U$ is,\footnote{It will be an interesting but non-trivial future project to find a similar rotating solution at $f_\pi\neq 0$. The restriction to $f_\pi=0$ for now allows us to find analytic solutions within the well-behaved sector of Einstein-Skyrme theory which is equivalent to  Einstein-Yang-Mills theory with a massless gauge field.}
\begin{equation}
U \,=\, n_1(r)  \mathds{1} + i n_3(r) \tau_z + i n_2(r) \left( \tau_x \cos(\phi - \omega t) + \tau_y \sin(\phi -\omega t) \right)\, , \label{eq:Rotatingansatz}
\end{equation}
The metric in this case is taken to be of the form,
\begin{equation}
    ds^2\,=\, -A(r)\,dt^2 + B(r) dr^2 + F(r) (dx^2 + dy^2) + 2\,H(r)\,dt\,d\phi + M(r)\,d\phi^2\, ,
\label{eq:metric}    
\end{equation}
where $\phi$ is a compact coordinate with range $\{0,2\,\pi\}$. Hence, the topology is $\mathbb{R}^{3,1}\times S^1$. We have used the scaling symmetry of the metric to set $L=1$ in this section in order to prevent cluttering of the equations. 
Imposing the constraint in eq.\ (\ref{eq:constraint}) and making the replacement
\be 
M(r)\,=\, \frac{Q(r) - H(r)}{\omega}\, ,
\ee
reduces the set of six independent Einstein equations to three equations in terms of five unknown functions.  One of the equations has the solution $Q(r)\,=\,F(r)$, and another of the three equations, after substituting for $Q(r)$, has the form
\be
B(r)\,=\, \frac{F\,'(r)^2}{4\,F(r)^2}\, ,
\ee
where $F\,'(r)\,=\,d F(r)/d r$. The one remaining equation has three unknown functions $\{ F(r), H(r), \chi(r) \}$ two of which can be arbitrarily chosen.  
A space-time which is asymptotically $AdS$ can be obtained by the choices
\begin{align}
F(r)&= a\,r^2 + b,\\
\chi(r)&= \arccos\left(\chi_1-\frac{\chi_2}{2\left(r^2-r_0^2\right)}\right)\, .\label{eq:Rotating_Skyrme_Solution}
\end{align}
In these expressions $a,\, b,$ and $r_0$ are constants.  
Although at this point the choice of Skyrmion is  arbitrary in our Einstein-Skyrme theory, the form chosen for $\chi(r)$ has been informed by our requirement of a mapping between our Einstein-Skyrme system and an Einstein-Meron-Yang-Mills theory as described in section~\ref{sec:merons}. 
In order for a horizon to exist and for the space-time to be asymptotically $AdS$, the constant $b$ must satisfy $b\,=\,-a\,r_0^2$. With these restrictions on the constants taken into account the metric functions can be expressed as
\begin{eqnarray} \label{eq:rotatingBraneSoln}
A(r)&=&\frac{\omega  \left(3 e^2 \left(r^2-r_0^2\right) \left(4 a \left(r^2-r_0^2\right)^2+c_2 r^2 \left(r^2-2 r_0^2\right)+4 c_1\right)-2 \chi _2^2 \omega \right)}{12 e^2 \left(r^2-r_0^2\right)^2} \, ,\nonumber \\
B(r)&=& \frac{r^2}{(r^2 - r_0^2)^2}\, ,\nonumber \\
F(r)&=& a\,(r^2 -r_0^2) \nonumber \, ,\\
H(r)&=&\frac{3 e^2 (r-r_0) (r+r_0) \left(c_2 r^2 \left(r^2-2 r_0^2\right)+4 c_1\right)-2 \chi _2^2 \omega }{12 e^2 \left(r^2-r_0^2\right)^2}\, , \nonumber\\
M(r)&=&\frac{2 \chi _2^2 \omega +3 e^2 (r-r_0) (r+r_0) \left(4 a \left(r^2-r_0^2\right)^2+2 c_2 r^2 r_0^2-c_2 r^4-4 c_1\right)}{12 e^2 \omega  \left(r^2-r_0^2\right)^2}\, .
\label{eq:rotating_Solution}
\end{eqnarray}
The dependence of these expressions on the constants $e,\, r_0,\,$ and $\omega$ shows how the addition of the Skyrme field $\chi(r)$ can affect the geometry of a brane. In addition the solution depends on 2 undetermined coefficients\footnote{Note that $F(r)$ has been chosen for convenience. This choice creates a horizon and simplifies the form of the solution.}: $c_1$ and $c_2$. 
The functions in eq.~\eqref{eq:rotatingBraneSoln} have the following leading order terms in the near boundary expansions ($r\to \infty$)
\begin{equation}
    A= \frac{1}{4} \omega  (4 a+c_2) r^2 \, ,\quad
    B= \frac{1}{r^2} \, ,\quad
    F= a\, r^2 \, ,\quad\
    H= \frac{c_2}{4}r^2 \, ,\quad
    M=  -\frac{-4 a+c_2}{4 \omega }  r^2 \, .
\end{equation}
A scaling transformation 
\begin{equation}
    \hat t = \frac{1}{2} \sqrt{\omega  (4 a+c_2)} \, t \, ,\quad
    \hat x_{1,2} = \sqrt{a} \, x_{1,2}\, , \quad
    \hat \phi = \frac{1}{2} \sqrt{-\frac{-4 a+c_2}{\omega }} \, \phi \, ,
\end{equation} 
leads to $AdS_5$ close to its standard form at large $r$
\begin{equation}
    ds^2 = r^2 \left (-d\hat t^2+d\hat x_1^2 + d\hat x_2^2 + d\hat\phi^2 + \frac{2 c_2 }{\sqrt{16 a^2-c_2^2}} d\hat t d\hat \phi \right)
    +\frac{dr^2}{r^2} \, .\label{eq:scaledADS}
\end{equation}
We would like the metric to approach an $AdS_5$ metric with one compact spatial coordinate $\hat \phi$ in the limit that $\omega\rightarrow 0$; thus we choose to identify $c_2=\omega$. 
As the notation implies, we think of $c_2$ as the angular velocity of the Skyrme field solution~\eqref{eq:Rotatingansatz}, which will turn out to be proportional to the angular velocity of the rotating black brane.  
Now we note that the line element reduces to $AdS_5$ in the limit of vanishing angular velocity. In addition for the coordinate transformation not to become complex $\omega$ must be in the range $4a>\omega>-4a$. 
As discussed later in section~\ref{sec:Energy_Momentum} the remaining coefficients $a$ and $c_1$ can be defined by the calculation of Komar integrals. The resulting expressions for these coefficients are given as,
\begin{subequations}
\begin{align}
    a&=\frac{ \omega }{4 \sqrt{1-(J/M)^2}}, \label{eq:ang_mom_a}\\
   c_1&= \frac{\omega}{4 M}  \left(\frac{2 G_5 J^2}{\sqrt{1-J^2/M^2}} \left(1-\sqrt{1-J^2/M^2}\right)+M r_0^4\right).
\end{align}
\end{subequations}
where $J,M$ are the angular momentum and mass of the space-time respectively\footnote{The  derivation can be found in section~\ref{sec:Energy_Momentum}.}. We can see that the coefficient $a$ plays a role reminiscent of the $a_{\text{MP}}$ parameter in more familiar rotating space-times such as Meyers-Perry solutions where $a_{\text{MP}}=J/M$.  

We have explicitly checked that given the identifications in \eref{eq:identify}  the Einstein-Skyrmion solution, \eref{eq:rotating_Solution} and \eref{eq:Rotating_Skyrme_Solution}, satisfies the Einstein-Yang-Mills equations of motion following from the action~\eref{eq:meronAction} for $\lambda=-1/2$. 

The rotating solution does not depend explicitly on $t$ or $\phi$ hence has the following Killing  vectors\footnote{We use the standard notation for Killing vectors with $w$ the vector itself and $w^{\mu}$ as its components. In particular for $w$ we have $w=w^{\mu}\partial_{\mu}=w^{\phi}\partial_{\phi}$ with $w^{\phi}=1$ and all other components vanishing.},
\begin{equation}
    k=\partial_t \hspace{1cm} w=\partial_{\phi}.
    \end{equation}
We then have a conserved quantity for a particle with four velocity $u^{\mu}$ associated with its angular momentum defined as $L_{\text{obs.}}=-u^{\mu}w_{\mu}$.
Consider now an observer with time-like four-velocity falling into the black brane with zero angular momentum,
\begin{align}
   L_{\text{obs.}}=0&=-u^{\mu}w_{\mu}=-u^{\mu}g_{\mu\nu}w^\nu=-u^tg_{t\phi}-u^{\phi}g_{\phi\phi}\\
 \Rightarrow   -\frac{g_{t\phi}}{g_{\phi\phi}}&=\frac{u^{\phi}}{u^t}=\frac{\frac{\exd \phi}{\exd \tau}}{\frac{\exd t}{\exd \tau}}=\frac{\exd\phi}{\exd t}=\Omega.\label{eq:ang_mom_derivation}
    \end{align}
Inserting the expressions for $g_{t\phi}$ and $g_{\phi\phi}$ in eq.\ (\ref{eq:ang_mom_derivation}) we obtain the angular velocity of the black brane, 
\begin{equation}
    \Omega= \frac{\sqrt{4 a-\omega } \left(2 \chi _2^2 \omega -3 e^2 \left(r^2-r_0^2\right) \left(r^2 \omega  \left(r^2-2 r_0^2\right)+4 c_1\right)\right)}{\sqrt{4 a+\omega } \left(2 \chi _2^2 \omega +3 e^2 \left(r^2-r_0^2\right) \left(4 a \left(r^2-r_0^2\right)^2+2 r^2 r_0^2 \omega -r^4 \omega -4 c_1\right)\right)}\, .
\end{equation}
Taking the limit of $r\rightarrow r_0$ we find at the horizon $r=r_0$ the angular velocity to be,
\begin{equation}
  \Omega_H=\frac{\sqrt{4 a-\omega}}{\sqrt{4 a+\omega}}.
\end{equation}
Taking the limit as $r\rightarrow \infty$ we find that there is also a non-zero angular velocity at the $AdS$-boundary,
\begin{equation}
    \Omega_{\infty}=\lim_{r\rightarrow\infty} \Omega =-\frac{\omega }{\sqrt{4 a-\omega } \sqrt{4 a+\omega }}\, .
\end{equation}
Looking back now to eq.\ (\ref{eq:scaledADS}) we can see that we can rewrite the scaled asymptotic form of the $AdS$ metric as,
\begin{equation}
     ds^2 = r^2 \left (-d\hat t^2+d\hat x_1^2 + d\hat x_2^2 + d\hat\phi^2 -2\Omega_{\infty} d\hat t d\hat \phi \right)
    +\frac{dr^2}{r^2} \, .
\end{equation}
We consider the relative angular velocity given by,
\begin{equation}
    \Omega_{T}=\Omega_H-\Omega_{\infty}=\frac{4 a}{\sqrt{4 a-\omega } \sqrt{4 a+\omega }},\label{eq:Ang_Vel_Thermodynamics}
\end{equation}
as the relevant angular momentum for thermodynamic relations. One can note that the angular velocity at the $AdS$-boundary is in the opposite direction of the angular velocity at the horizon. For the angular velocity at the horizon to remain real-valued,   $\omega>0$. This is consistent with the bound from the entropy density as shown below. 
 
The entropy of the space-time associated with the horizon area can be computed as done in the previous section via $S=A/4G_5$. Computing the area $A$,
\begin{equation}
    A=\int \sqrt{\text{det}(g_{ij}(r_h))}=\sqrt{\frac{2}{3}} \sqrt{\frac{ \omega }{ (4 a-\omega)}}\frac{\chi_2}{e} \left(2\pi\int \exd^2 x\right)=\sqrt{\frac{2}{3}} \sqrt{\frac{ \omega }{ (4 a-\omega)}}\frac{\chi_2}{e} V,
\end{equation}
where $V=2\pi\int \exd^2x$ is an infinite Euclidean volume. We then find the entropy density to be given by,
\begin{equation}\label{eq:sRotating}
    s=S/V=\sqrt{\frac{2}{3}} \sqrt{\frac{ \omega }{ (4 a-\omega)}}\frac{\chi_2}{e}\, .
\end{equation}
As in the static case, the horizon area is directly proportional to the Skyrme parameter $\chi_2$. Furthermore as $\omega$ is bounded by $|\omega|<4a$ for the entropy to remain positive, the bound on $\omega$ is sharpened to $4a>\omega>0$. This is consistent with the bound found from the angular velocity at the horizon. As the Skyrmion angular velocity increases, the entropy density begins at zero and rapidly increases as $\omega$ approaches $4a$. 

We calculate the temperature via the Killing vector $v^{\mu}=k^{\mu}+\Omega_H w^{\mu}$, where we utilize the formula,
\begin{equation}
    \kappa^2=\frac{-1}{2}(\nabla_{\mu}v_{\nu})(\nabla^{\mu}v^{\nu})
\end{equation}
The Killing vector is given by $v^{\mu}=(1,0,0,\Omega_H,0)$, and a simple calculation gives the temperature for the rotating Skyrmion solution\footnote{Coordinates are given by $x^{\mu}=(t,x,y,\phi,r)$},
\begin{equation}
   \kappa =0 \quad\rightarrow\quad T=0\, .
\end{equation}
From this we see our solution represents an extremal rotating black brane solution. The finite entropy density at vanishing Hawking temperature is reminiscent of the finite entropy of five-dimensional rotating black holes~\cite{Lu:2008jk,Myers:1986un,Hawking:1998kw,Hawking:1999dp}. Of course, a non-vanishing entropy density is obtained also for extremal Reissner-Nordstr\"om black holes.

Unlike the previous section this is a zero temperature solution rendering the heat capacity not well-defined. For the purpose of a stability analysis~\cite{Chamblin:1999tk}, we can still use the ``angular momentum susceptiblity'' of the angular momentum $J$ to the angular velocity $\Omega_T$. We proceed by starting with the internal energy as defined in~\cite{Papadimitriou:2005ii},\footnote{It is curious that in this expression the charge and chemical potential term does not appear as it would for a simple Reissner-Nordstr\"om black hole. Rewriting the Skyrmion matter action into a Yang-Mills action as discussed in section~\ref{sec:merons}, one may have naively expected that the Skyrme matter would be equivalently creating a charged black hole. This is one instance in which it becomes clear that the thermodynamic relations for the rotating Skyrme-$AdS_5$ black brane solutions are not simply those of Einstein gravity plus a Yang-Mills gauge field. Thus, these are not simply Reissner-Nordstr\"om solutions in disguise.} 
\begin{equation}
    d M=T\exd S+\Omega_T \exd J
\end{equation}
from which we perform a Legendre transformation,
\begin{equation}
    F_J=M-J\Omega_T \quad\rightarrow \quad \exd F_J=T\exd S- J\exd \Omega_T\, .
\end{equation}
As shown above the temperature is zero in our system and hence,
\begin{equation}
    \exd F_J=- J\exd \Omega_T\, ,
\end{equation}
from which we find using the internal energy,
\begin{equation}
    \frac{\partial F_J}{\partial \Omega_T}=-J\, .
\end{equation}
We can define the angular momentum susceptiblity as the second variation of $F_J$ with respect to the angular velocity giving,
\begin{equation}\label{eq:cOmega}
    \chi_{{}_\Omega}= \frac{\partial^2 F_J}{\partial \Omega_T^2}= \frac{\partial J}{\partial \Omega_T}=\frac{M}{\Omega_T^2}\geq 0 \, ,
\end{equation}
where $J$ will be discussed in detail in section~\ref{sec:holoSkyrmionsMerons}. Here the angular momentum $J$ and the angular velocity $\Omega_T$ are conjugate quantities. 
The inequality in~\eqref{eq:cOmega} is the condition for stability against rotational fluctuations, which holds for our rotating solution. 

In order to ensure stability against charge fluctuations, we now consider the electric susceptibility with respect to the charge as defined in the previous section,
\begin{equation}
   (\chi_\phi)^a_b=\frac{\partial Q^a}{\partial \mu^b}\, .
\end{equation}
Using the expressions given in eq.\ (\ref{eq:Ext_Gauge_Rotating_t}) and eq.\ (\ref{eq:ROT_current_t}) we find, 
\begin{equation}
(\chi_\phi)^a_b=
\begin{pmatrix}
 \frac{\left(1-2 \chi _1^2\right) \chi _2}{2 \chi _1 \left(\chi _1^2-1\right)} & 0 & 0 \\
 0 & \frac{\left(1-2 \chi _1^2\right) \chi _2}{2 \chi _1 \left(\chi _1^2-1\right)} & 0 \\
 0 & 0 & \frac{\chi _1 \chi _2}{1-\chi _1^2} \\
\end{pmatrix} \, .
\end{equation}
There is a direct similarity to susceptibility matrix as displayed in eq.\ (\ref{eq:Static_sus}). This is due to the similarity in the ansatz of the Skyrme field given in eq.\ (\ref{eq:Sansatz}) and eq.\ (\ref{eq:Rotatingansatz}). The difference in sign between the two susceptibility tensors comes from a difference in sign for $\chi_1$ in the solution to the Einstein-Skyrme field equations. We find that,
\begin{equation}
   (\chi_\phi)^a_b>0 \quad \text{if}\quad
  \begin{cases}
        1>\chi_1\geq1/\sqrt{2} & \text{\&} \quad\chi_2>0 \\
        -1<\chi_1\leq-1/\sqrt{2} &\text{ \&} \quad \chi_2<0
    \end{cases}\, .
\end{equation}
Considering our theory with these parameter restrictions ensures that our solution is stable against isospin charge fluctuations. 

\subsection{Topological Skyrme-$AdS_5$ black holes} 
\label{sec:topologicalMeronSolutions}
The analytic solutions presented in the previous subsections are topologically trivial in that they have vanishing topological charge  $\mathpzc{q}$, according to eq.~\eqref{eq:instantonCharge}. However, an analytic meron solution to the
massless $SU(2)$ Einstein-Yang-Mills theory with non-trivial topological charge can be computed when we take the definition of the winding number seriously. This quantity counts the number of times the mapping wraps the internal $SU(2)$ group manifold. In general this can be written as,
\begin{equation}
    U=e^{i n \chi v^i \tau^i},\label{eq:top_Ansatz}
\end{equation}
where $v^iv^i=1$ and $n\in \mathbb{Z}$. The integer $n$ quantifies how many times our mapping wraps the internal $S^3$. This expression can be written in the form of our original ansatz as,
\begin{equation}
    U=\cos(n \chi) \mathds{1}+i\sin(n\chi)v^i\tau^i.\label{eq:Gen_Topological_Ansatz}
\end{equation}
We choose to work with coordinates $(t,\psi,\theta,\phi,r)$
where $\psi\in (0,\pi)$, $\theta\in (0,\pi)$, $\phi\in (0,2\pi)$, $t$ is our temporal coordinate and $r$ is the bulk $AdS$ radial coordinate. It is particularly useful to choose the form of the unit vector $v$ as,
\begin{equation}
    v=(\cos (\theta ),\sin (\theta ) \cos (\phi ),\sin (\theta ) \sin (\phi )),\label{eq:unit}
\end{equation}
along with setting  $\chi=\psi$. For $n=1$ this results in a standard mapping of the unit three sphere into the $SU(2)$ gauge manifold. We take as an asatz for the space-time metric,
\begin{equation}
    ds^2=\frac{1}{A(r)}\exd r^2-A(r)\exd t^2+r^2\left(h_1(\psi)\exd \psi^2+h_2(\psi)(\exd\theta^2+\sin(\theta)^2\exd\phi^2)\right).
\end{equation}
The Einstein-Skyrme equations at vanishing pion coupling, $f_\pi=0$, result in three independent equations for $h_1,h_2$ and $A$. The solutions of these equations are,
  \begin{equation}
    h_1(\psi)=n^2,\quad h_2(\psi)=\sin(n\psi),\quad A(r)=\frac{1}{e^2 r^2}+\frac{\log (r)}{e^2 r^2}+\frac{r^2}{L^2}-\frac{m_t}{r^2}+1 \label{eq:Topo_non_trivial_Metric}\, .
\end{equation} 
Using the solutions eq.\ (\ref{eq:Topo_non_trivial_Metric}) along with $K_{\mu}$ created from eq.\ (\ref{eq:Gen_Topological_Ansatz}) and eq.\ (\ref{eq:unit}) the Skyrme equations of motion, along with the Yang-Mills equations of motion with $\lambda=-1/2$, are trivially satisfied. 
As explained in section~\ref{sec:merons} the Skyrmion theory can be expressed equivalently as an $SU(2)$ gauge theory. The logarithmic term in \eqref{eq:Topo_non_trivial_Metric} 
usually signals a conformal anomaly. 
However, the case at hand is more subtle and will be discussed in section~\ref{sec:holoSkyrmionsMerons}.

The last three terms in $A(r)$ are the global $AdS$ blackening factor, see e.g.~\cite{GGdual,Karch:2006bv}. This can be expected due to the boundary geometry being $\mathbb{R}\times S^3$. It is interesting to note that the terms with factors of the Skyrme model parameter $e$ are new and deform this solution away from global $AdS_5$.
It may appear that a competition between the term $1/(e^2 r^2)$ with the mass and logarithmic term could make the horizon radius zero if $e$ is chosen properly. However, this is not the case. There is only a single simultaneous limit for which the horizon location goes to zero corresponding to $m_t\rightarrow 0, e\rightarrow \infty$. The single limit of $e\rightarrow \infty$ leads us back to the global Schwarzschild black hole.

\begin{figure}[htb]
    \begin{subfigure}[b]{0.5\textwidth}
    \includegraphics[width=80mm,scale=0.5]{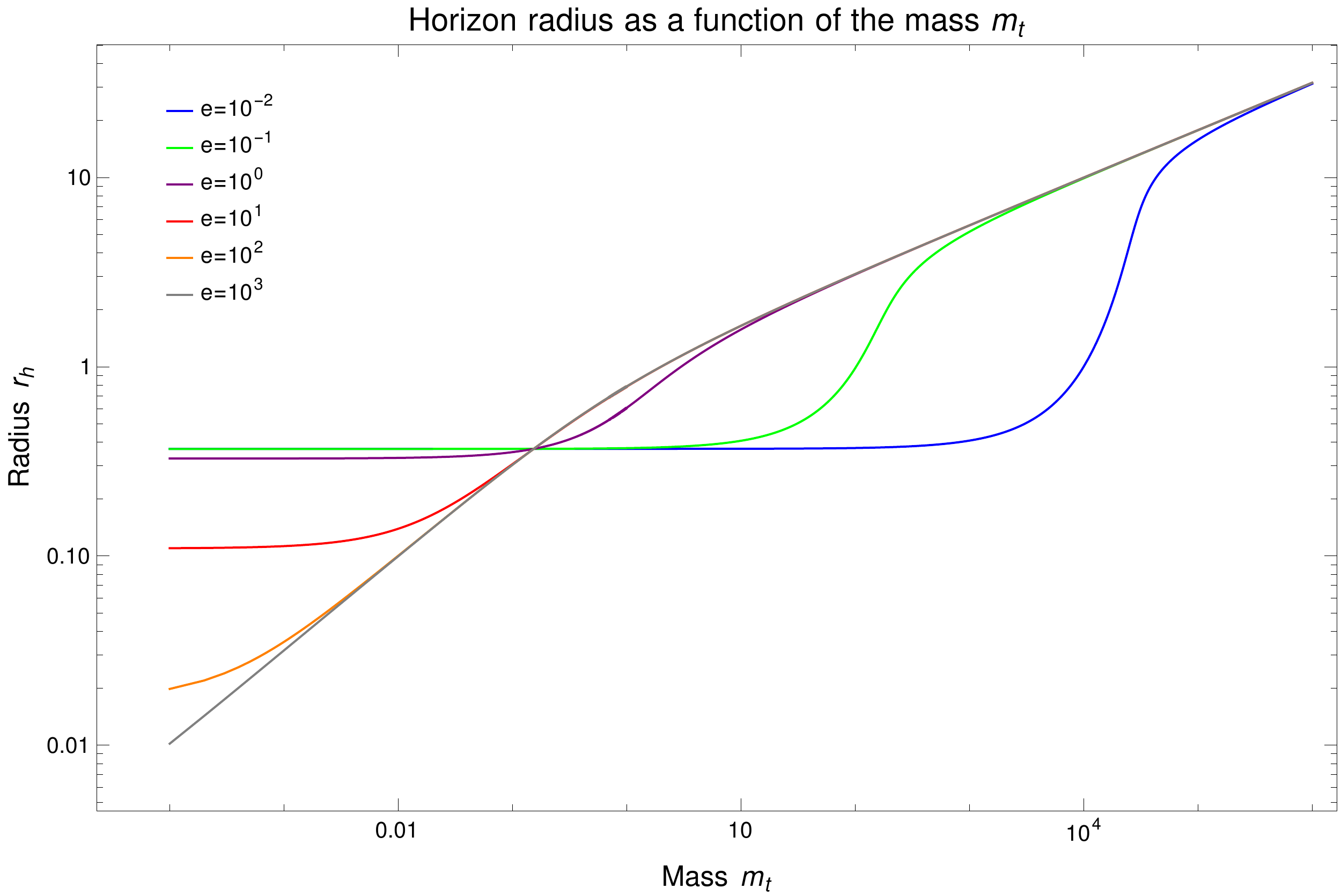}
    \end{subfigure}
    \begin{subfigure}[b]{0.5\textwidth}
    \includegraphics[width=80mm,scale=0.5]{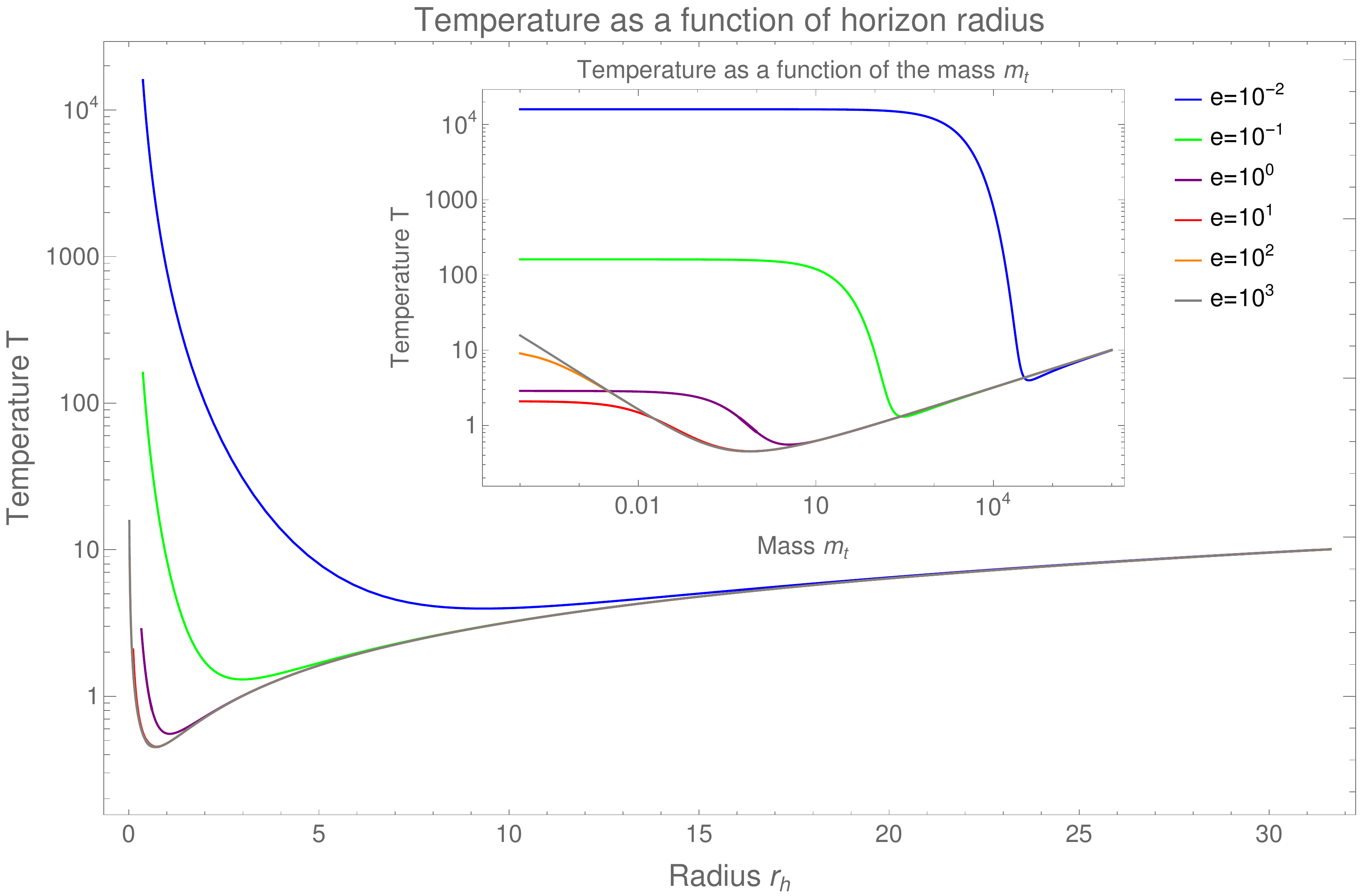}
    \end{subfigure}
    \caption{
    Gravitational data for the topologically non-trivial solution. 
    \textit{Left:} The horizon radius, $r_h$ as a function of the mass $m_t$, is displayed on a log-log plot. \textit{Right:} The temperature $T$ as a function of the horizon radius $r_h$ is displayed on a log plot. The inset graphic displays the temperature $T$ as a function of the mass. All images are displayed with the Skyrme model parameter scanned over a window spanning six orders of magnitude with $e=(10^{-2},10^{-1},10^{0},10^{1},10^{2},10^{3})$. Between the left and right image the colors correspond to the same value of the Skyrme model parameter. 
    \label{fig:Topologically_Non_Trivial}}
\end{figure}

Although our metric is expressed in a closed form~\eqref{eq:Topo_non_trivial_Metric}, the logarithm in the blackening factor does not allow an explicit expression for the location of the horizon radius or the temperature. Hence, we solve the resulting transcendental equation numerically. The resulting horizon radius as a function of the black hole mass, $r_h(m_t)$, and the temperature as a function of the horizon radius, $T(r_h)$ (and mass $T(m_t)$) are displayed in figure~\ref{fig:Topologically_Non_Trivial}. 
At small Skyrme-coupling, $e\gg 1$, the behavior of the temperature of a global $AdS$ black hole is restored. 
There exists a well-known minimum temperature in the global $AdS_5$ black hole, $T_{\text{min}}=\sqrt{2}/(\pi L)$, which is shifted to larger temperatures (at larger radii) as $e$ decreases from infinity towards zero.  
As usual, for one particular temperature, there are (at least) two horizon radii. This indicates that there are two black hole solutions with the same temperature, one small black hole and one large black hole. As the influence of the Skyrmion is increased, lowering $e$, the minimum in the curve shifts to larger radii and eventually disappears. 
Plugging the value for $r_h$ obtained by Mathematica's FindRoot~\cite{Mathematica} back into the blackening factor reveals that the residuals are $O(10^{-11})$.   
That minimum is pushed to larger horizon radii and disappears as $e\to 0$. We have checked that the transition from the lowest curve ($e=10^3$) in the right of figure~\ref{fig:Topologically_Non_Trivial} to highest curve ($e=10^{-2}$) is smooth as we increase the Skyrme model parameter $e$.
This behavior results from a competition between the Skyrmion with itself and with the black hole geometry, showing in~\eqref{eq:Topo_non_trivial_Metric}, as mentioned in the previous paragraph.

If we consider the limit as $e\rightarrow 0$, this corresponds to an infinite contribution of the Skyrmion in the bulk geometry, we note that the black hole of this configuration has a finite horizon radius determined by
\begin{equation}
A(r_h)=0=\frac{1}{e^2 r_h^2}+\frac{\log (r_h)}{e^2 r_h^2}+\frac{r_h^2}{L^2}-\frac{m_t}{r_h^2}+1 \hookrightarrow \frac{1}{ r_h^2}+\frac{\log (r_h)}{r_h^2}+\frac{e^2r_h^2}{L^2}-\frac{e^2\,m_t}{r_h^2}+e^2=0 \, .
\end{equation}
Now taking the limit as $e\rightarrow 0$ we find,
\begin{equation}
    \frac{1}{ r_h^2}+\frac{\log (r_h)}{r_h^2}=0 \rightarrow r_h=\text{exp({-1})}\approx 0.368\, .
\end{equation}
The black hole attains a finite horizon radius as $e\rightarrow 0$, and as this occurs $T\rightarrow \infty$. 
The entropy turns out to be quantized, 
\begin{equation}
    S = 
    \frac{2\pi^2 n}{4\,G_5} r_h^3\, , \quad n\in \mathbb{N}  \, .
\end{equation}
This is in line with the fact that the energy of this solution gains a quantum of energy each time the winding number $n$ of the topological Skyrmion solution is increased. 

The well-known Hawking-Page transition~\cite{hawking1982} from thermal $AdS$ to an $AdS$ black hole space-time occurs in our solution in the limit $e\rightarrow \infty$. In this limit the critical horizon value is $r_h=L$, which determines the critical temperature $T_c=3/(2\pi L)$. By computing the free energy associated with our topological Skyrme black hole solution and comparing it with that of thermal $AdS$ we discern the effect the Skyrmion has on the transition. We compute the renormalized on-shell action as,
\begin{equation}
 S_{ren}=  \lim_{\epsilon\rightarrow 0}\left(\int \exd^4 x\left(\int_{r_h}^{1/\epsilon}\exd r\sqrt{-g}\left(\frac{(R-2\Lambda)}{16\pi G_5}+\frac{\tr(F^2)}{16\pi \gamma^2}\right)   \right)+S_{ct}\right),
\end{equation}
where $S_{ct}$ is the counter-term action. This action is given as the usual Gibbons-Hawking-York boundary term needed to make the variational problem well defined plus additional contributions needed to cancel both $1/\epsilon^n$ and $\log(\epsilon)$ divergences~\cite{Taylor:2000xw},
\begin{equation}
    S_{ct}=\frac{1}{8\pi G_5}\int\exd^4 x\sqrt{\gamma}\left( K -\frac{1}{2L}\left(2(1-d)-\frac{L^2}{d-2}R(\gamma)  \right)\right)+ \frac{L}{16\pi\gamma^2}\log(\epsilon)\int\exd^4 x\sqrt{\gamma_0}\tr(F_0^2)\, ,
\end{equation}
where $K$ is the trace of the extrinsic curvature, 
$\gamma$ is the induced metric on a constant $r=1/\epsilon$ hypersurface, $\gamma_0$ is the metric of the dual field theory and $F_0$ is the external field strength of the gauge field $A$ in the dual theory. Applying this formula we find the following difference in free energy between our topological $AdS$ black hole and thermal $AdS$, 
\begin{equation}\label{eq:DeltaFHawkingPage}
 \Delta F = F_{\text{BH}}-F_{\text{thermal}}= \frac{\pi ^2 \beta  n r_h^2 }{\kappa}\left(-\frac{\tilde{e}^2 L^4 \left(\frac{3}{n}-3\right)+4 \tilde{e}^2 r_h^4+4 L^2}{4 \tilde{e}^2 L^2 r_h^2}+\frac{3 \kappa  \log (r_h)}{\tilde{e}^2 r_h^2}+1\right)\, .
\end{equation}
Notice, as stated in the previous paragraph, if we take the limit $e\rightarrow\infty$, the horizon radius asymptotes to the horizon radius of the standard global $AdS_5$ black hole, $r_h \to r_h(e\to \infty)$, and leads to the free energy difference,   
\begin{equation}
    \lim_{e\rightarrow\infty}\frac{\kappa }{\pi ^2 \beta r_h^2}\Delta F=\left(1-\frac{r_h^2}{L^2}\right) \, .
\end{equation}
Hence, we restore the standard Hawking-Page transition temperature defined by $r_h=L$. However, if $\infty>e>0$ and we consider e.g.~$n=1$, then the $\text{log}(r_h)$-term and the $1/(e^2 r_h^2)$-terms compete with $r_h^2/L^2$.

\begin{figure}[htb]
\begin{center}
    \includegraphics[width=14cm]{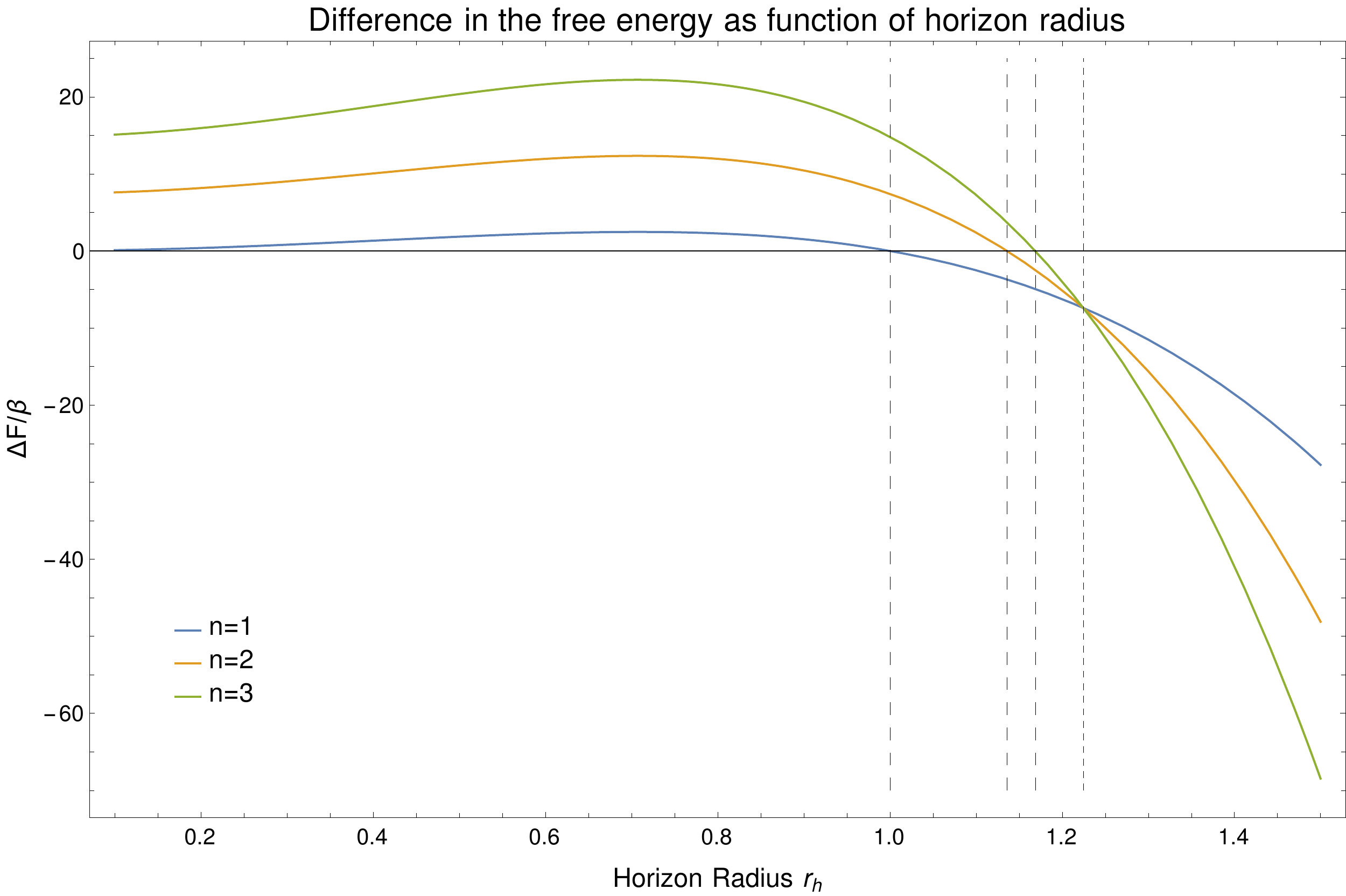}
    \end{center}
    \caption{
    \textit{Skyrmion shift of the Hawking-Page transition:} Difference in the free energy $\Delta F$ per inverse temperature $\beta$ between the topological Skyrme-$AdS_5$ black hole and thermal $AdS$ space-time as a function of the horizon radius $r_h$. The three curves correspond to different winding numbers, \textcolor{blue}{$n=1$} (blue), \textcolor{orange}{$n=2$} (orange) and \textcolor{green}{$n=3$} (green). The black lines with larger dashing represent the intersection of the difference in free energy with zero indicating the point at which the topological Skyrme-$AdS_5$ black hole has the same free energy as thermal $AdS$. These intersections occur at $r_h=1$ for $n=1$, $r_h=1.13$ for $n=2$ and $r_h=1.16$ for $n=3$. The black line with fine dashing represents the intersection of all three curves at $r_h=1.22$ indicating all three of the differences of the free energies take on the same value. Hence none of the three different winding numbers displayed is thermodynamically preferred over the others. All curves displayed above are calculated for $\tilde{e}=10^6$.
    A discussion of the focal point at $r_h\approx 1.22$ may be found at the end of section~\ref{sec:topologicalMeronSolutions}.\label{fig:Topologically_Non_Trivial_Free_Energy}}
\end{figure}

The free energy difference~\eqref{eq:DeltaFHawkingPage} is displayed in figure~\ref{fig:Topologically_Non_Trivial_Free_Energy} as a function of the horizon radius. These results are generated at a large value of the Skyrme coupling $\tilde e=10^6$, i.e.~with a weak influence of the Skyrmion on the geometry. The lowest (blue) curve is virtually identical to the standard free energy for a thermal $AdS_5$ to global $AdS_5$ black hole transition at a critical temperature fixed by $r_h=L$. Increasing the winding number of our solution, this critical horizon radius increases by a $\Delta r_h$ which is a discrete function of $n$, shifting the Hawking-Page phase transition. This trend continues for larger $n>2$ as well. Below the transition, the configurations with larger $n$ have a larger free energy. However, there exists a ``focal point'' at a certain $r_h$ where all the free energies intersect. That means there is no energetically favored topology at that horizon radius (or corresponding temperature). Above that horizon radius, larger winding numbers are energetically preferred. As can be seen from \eqref{eq:DeltaFHawkingPage}, the contribution from the thermal $AdS$ free energy, $F_{thermal}$, (the $3/n$-term) is independent of the winding number $n$.  The remaining terms in the expression are proportional to $n$, which can be factored out, leaving an expression which is a polynomial in $r_h$.  The polynomial vanishes at $ r_h\,=\,1.22$ for $\tilde{e}\,=\,10^6$ and $\kappa\,=\,1$, leaving $F_{thermal}$ as the only contribution to $\Delta F$. Thus the parameters of the Skyrmion field can be chosen such that the free energy of the topological black hole is zero for all winding numbers.
In other words, in figure~\ref{fig:Topologically_Non_Trivial_Free_Energy} at $r_h=1.22$ there is only one contribution to $\Delta F$ which is independent of $n$, hence the curves for different $n$ intersect in the focal point at that horizon value.  These are rather interesting features of our solutions, which we will discuss elsewhere~\cite{FutureCartwright}. 

\section{Holographic interpretation}
\label{sec:holoSkyrmionsMerons}
In this section, we provide an interpretation of our solutions from section~\ref{sec:blackBraneSolutions} in the holographically dual field theory. 
In section~\ref{sec:merons}, we have shown that one may think of the Skyrmion solutions, see section~\ref{sec:Skyrmions}, as being equivalent to certain $SU(2)$ gauge field configurations known as merons. In particular this applies to the explicit Einstein-Skyrme solutions we found: a static black brane in section~\ref{sec:staticBBs}, a rotating black brane in section~\ref{sec:rotatingBBs}, and a topologically non-trivial black hole in section~\ref{sec:topologicalMeronSolutions}. 
Hence, we can interpret these gravity solutions as being dual to a field theory containing a current $J^\mu$ carrying an $SU(2)$ charge. 

When evaluated at vanishing Proca mass, $m=0$, the meron action~\eqref{eq:meronAction} coincides with the Einstein-Yang-Mills action with an $SU(2)$ gauge field, $A_\mu$. This is a well-known consistent truncation of the bosonic part of minimal gauged type IIB supergravity in five dimensions. As mentioned in the introduction, we choose to set the Chern-Simons term to zero in order to simplify the analysis, see footnote~\ref{foo:ChernSimonsNeglected}. 
The massless case is considered throughout this paper with the exception of section~\ref{sec:staticBBs}, where we demonstrate the validity of our solutions within the Skyrmion theory with nonzero kinetic term, $f_\pi\neq 0$ which requires a nonzero Proca mass term, $m\neq 0$, as seen from~\eqref{eq:identify}.  
While our analysis does not require it here, we note that such a mass can be generated within type IIB supergravity through spontaneous symmetry breaking in the bulk, for example demonstrated by breaking a $U(1)$-subgroup of the R-symmetry in~\cite{Klebanov:2002gr}. There, the dual field theory acquires an anomaly breaking the global $U(1)$ R-symmetry. For our $SU(2)$ R-symmetry subgroup we would expect a similar mechanism, however, we leave this study for the future and now focus on massless gauge fields in the bulk. 
In other words, the massless meron theory corresponds to $\mathcal{N}=4$ Super-Yang-Mills theory~(SYM) in flat Minkowski space coupled to an external $SU(2)$ gauge field $F$ associated with an $SU(2)$ subgroup of the $SU(4)$ R-symmetry of the $\mathcal{N}=4$ SYM theory, see e.g.~\cite{Son:2006em,Behrndt:1998jd,Cvetic:1999ne,Gubser:1998jb,Chamblin:1999tk}.
\footnote{
Including the proper Chern-Simons term may have interesting effects on the topological properties of the solutions, see e.g.~\cite{Cartwright:2020qov,Ammon:2016szz}, and is subject to future study~\cite{FutureCartwright}.}

Alternatively, a term similar to the $F^2$-term in the action~\eqref{eq:meronAction} could arise as the leading order (in gauge field strength) term of the truncated probe D-brane action. An example is the D3/D7-brane intersection~\cite{Karch:2002sh} with two coincident $D7$-branes. This introduces two types ({\it flavors}) of fundamental fermions ({\it quarks}) into the $\mathcal{N}=4$ SYM theory, breaking the supersymmetry to $\mathcal{N}=2$. In this case, the current $J^\mu$ should be interpreted as a flavor current, associated with an $SU(2)$ symmetry, and the associated conserved charge may be interpreted as the isospin charge in QCD~\cite{Kaminski:2008ai,Erdmenger:2007ap,Erdmenger:2007ja,Erdmenger:2008yj,Ammon:2008fc}. We should note here, that in general the determinant of the metric of such a truncated D-brane action can be more complicated than simply the determinant of the underlying asymptotically $AdS$ space-time metric. That is because the relevant metric induced on the worldvolume of the D7-branes can contain contributions from the gauge field, see e.g.~\cite{Myers:2008cj,Ammon:2008fc,Ammon:2009fe}. 

Since the solutions in section~\ref{sec:blackBraneSolutions} are black branes/holes, generally the dual state is a strongly coupled quantum many-body system at non-zero temperature. For example, one may imagine the quark-gluon-plasma generated in heavy-ion collisions, or a strongly-correlated electron system. 
The Skyrme hair of the branes is dual to the gauge field sourcing the $SU(2)$ current $J^\mu$. That current is restricted in that it is dual only to merons, i.e.~pure gauge solutions. 

We stress again, that none of our results requires a non-zero meron mass. However, allowing a non-zero meron mass, the current $J^\mu$ is not conserved. 
Nevertheless, the divergence of the current is precisely known and proportional to the squared meron mass, 
\begin{equation}\label{eq:divergenceJ}
    \partial_jJ^j = c_0 m^2 \partial_i A_{(0)}^i \, ,
\end{equation}
where $A_{(0)}=\lim\limits_{r\to r_{\text{bdy}}}A$ is the bulk gauge field evaluated at the $AdS$-boundary, $m$ is the meron mass, $c_0$ is a constant depending on the black brane solution, i.e.~on the microscopic properties of the dual field theory, and $i,j=0,\,1,\,2,\,3$ are indices parametrizing the boundary coordinates. Note that through the meron mass the non-zero divergence in eq.~\eqref{eq:divergenceJ} is related to the operator dimension $\Delta$ of the operators added to the $\mathcal{N}=4$ SYM mentioned before, which break the $SU(2)$ and conformal symmetry. 
We now discuss the three solutions presented in section~\ref{sec:blackBraneSolutions}.

\subsection{Currents and external fields}
In the absence of the meron mass term in the meron action~\eqref{eq:meronAction}, i.e. $f_{\pi}=0$, the current $J$ is in principle conserved. It is computed by expanding the meron field $A_\mu$ near the $AdS$-boundary, and extracting the coefficient of the normalizable mode. 
We recall that our gauge field $A_{\mu}=-1/2 K_{\mu}$ has operator dimension one.\footnote{The operator dimension $\Delta$ for a $p$-form field is related to the mass $m$ of that field by $m^2 L^2 = (\Delta - p)(\Delta+p-d)$ with $d=4$, $p=1$, and $m=0$ for the gauge field $A_\mu$. This yields $\Delta_\pm = 1,\, 3$, determining the near-boundary exponents of the normalizable and non-normalizable modes of $A_\mu$.} 
Hence, we can expand $A_{\mu}$ near the $AdS$-boundary as,
\begin{equation}
 A_{\mu}=  \frac{-1}{2} A^{(0),a}_{\mu}\tau_a -\frac{1}{2r^2}A^{(2),a}_{\mu}\tau_a+ O(r^{-3})\, ,
\end{equation}
revealing the normalizable and the non-normalizable mode. 
The standard $AdS$/CFT dictionary tells us we can identify these with the vacuum expectation value of the dual operator and its source, respectively,
\begin{subequations}
\begin{align}
    \vev{J_{i}}&=\lim_{r\rightarrow \infty}r^2A_i=\frac{-1}{2} A^{(2),a}_{i}\tau_a \, ,\label{eq:dict_vev}\\
    A^{ext}_{i}&=\lim_{r\rightarrow \infty}A_{i}=\frac{-1}{2} A^{(0),a}_{i}\tau_a \, \label{eq:dict_sour}.
\end{align}
\end{subequations}
In other words, here $A^{ext}$ is the externally applied $SU(2)$ gauge field configuration and $\vev{J}$ is the vacuum expectation value of the global $SU(2)$ current. We can identify an $SU(2)$ chemical potential as the time component of the external gauge field $\mu=A^{ext}_{t}$.

\textbf{Static solution:} For this case we find the only non-zero component of the bulk gauge field to be $A_t$ with,
\begin{equation}
    \mu^a = A^{(0),a}_t=2\left(2 \chi_1 \sqrt{1-\chi_1^2} \omega  \sin (t \omega ),2 \chi_1 \sqrt{1-\chi_1^2} \omega  \cos (t \omega ),-2 \left(\chi_1^2-1\right) \omega \right)\label{eq:Ext_Gauge_static} \, ,
\end{equation}
and 
\begin{equation}\label{eq:vevStatic}
  \braket{J^{a}_t}=  A^{(2),a}_t=-\left(\frac{\left(2 \chi_1^2-1\right) \chi_2 \omega  \sin (t \omega )}{\sqrt{1-\chi_1^2}},\frac{\left(2 \chi_1^2-1\right) \chi_2 \omega  \cos (t \omega )}{\sqrt{1-\chi_1^2}} ,2 \chi_1 \chi_2 \omega  \right) \, ,
\end{equation}
It is interesting to check the divergence of this current via $\partial_{i}\braket{J^{i,a}}$ which gives,
\begin{equation}\label{eq:divJStatic}
    \partial_{i}\braket{J^{i,a}}=\partial_t\braket{J^{t,a}}=\frac{\chi_2(2\chi_1^2-1)}{2\sqrt{1-\chi_1^2}}\omega^2 \left( \cos(\omega t),  -\sin(\omega t), 0\right),
\end{equation}
where we see that we have a non-conserved global charge associated to our $SU(2)$ gauge field. 

A few observations are in order. First, a simple interpretation of the non-conservation is due to the oscillating chemical potential inducing an oscillating charge density. The charge density $\vev{J^{a}_t}$ and chemical potential $\mu^a$ are always in phase while the divergence of the current $\vev{J}$ and the chemical potential $\mu^a$ are always out of phase. By suitable adjustment of the parameters $\chi_1$ or $\chi_2$ we may have a conserved current $\vev{J}$ ($\chi_1=\pm 1/\sqrt{2}$). We note that out of this two-parameter family the values $|\chi_1|=1,\, \chi_2=0$ are pathological. 
However if we take $\omega=0$ all sources $\mu^a$ and vacuum expectation values $\vev{J}$ vanish. Furthermore it is interesting to note that although the charges are non-conserved, the current averaged over one cycle of period $2\pi/\omega$ is conserved. We note in passing that although there is a non-zero external $SU(2)$ gauge field in the dual field theory, the field strength associated with this gauge potential is trivially zero, i.\ e.\ $F^a_{ij}=0$. 

In section~\ref{sec:blackBraneSolutions} we showed that our solution for the static black brane was smoothly connected to the solution with a mass term present for the meron gauge field. Utilizing the equations given in eq.\ (\ref{eq:dict_vev}) and eq.\ (\ref{eq:dict_sour}) for the massive meron solution leads to a vanishing value for the vacuum expectation value of the dual current $J$ and the external gauge field $A^{ext}$.

\begin{figure}
    \centering
    \includegraphics[width=12cm]{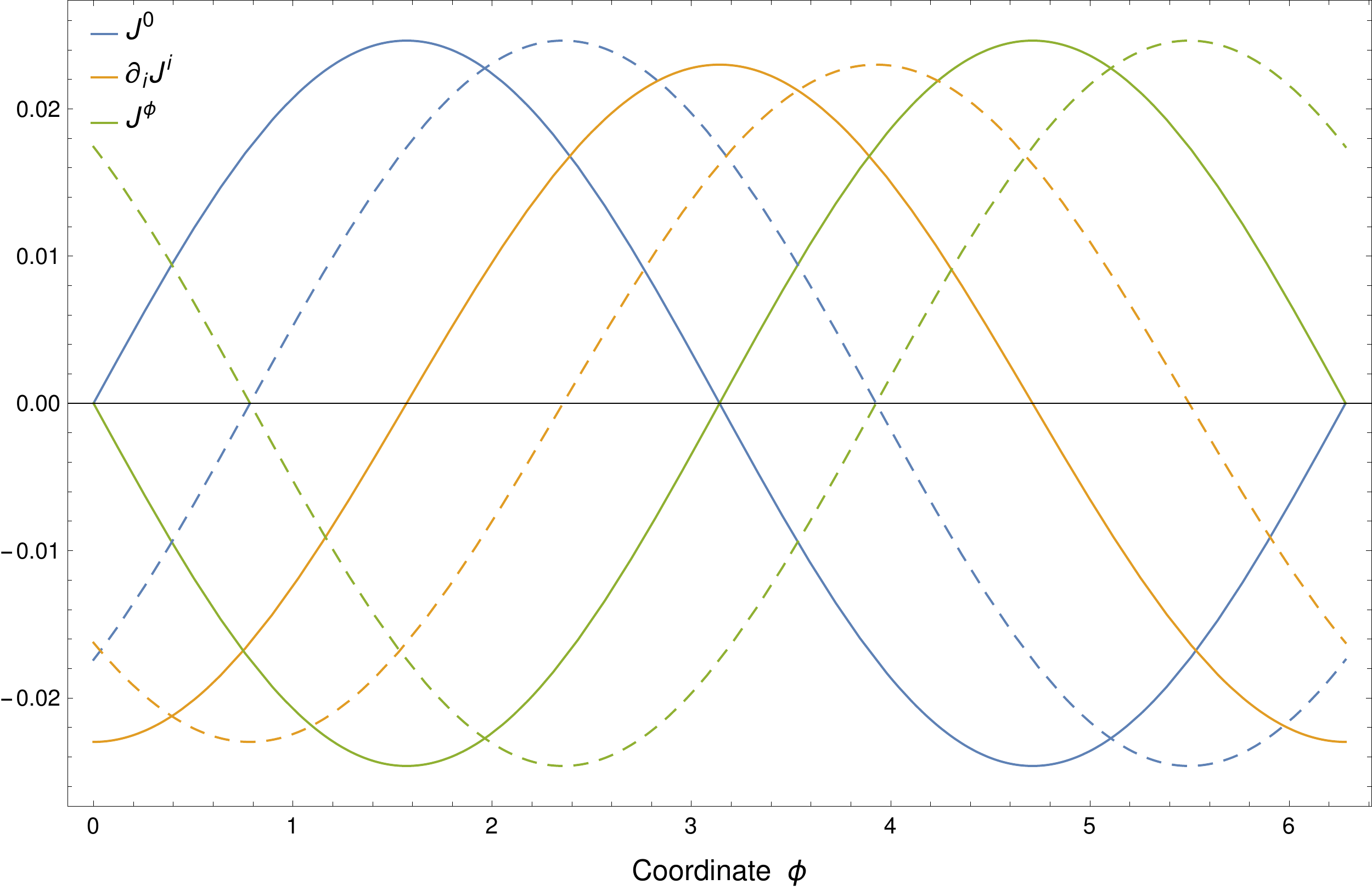}
    \caption{The $t^1$ component of the charge, current and divergence ``waves'' in the dual theory are displayed. The solid lines are displayed at $t=0$ while the dashed lines are displayed at $t=\pi/4\omega$. For this image we make the choice $\omega=1,\chi_1=0.1,\chi_2=0.1$ and $a=4$. As discussed in the text we can see that the 
    \label{fig:currentwaves}}
\end{figure}

\textbf{Rotating solution:}
In the case of the rotating black brane solution we find non-zero components of both $\braket{J_t^a}$ and $\braket{J_\phi^a}$. The sources of these operators are,
\begin{align}
     \mu^a = A^{(0),a}_{t}&=\left( -  \chi_1 \sqrt{1-\chi_1^2} \omega  \sin (\phi -t  \omega ), \chi_1 \sqrt{1-\chi_1^2} \omega  \cos (\phi -t  \omega ), \left(\chi_1^2-1\right) \omega     \right)\label{eq:Ext_Gauge_Rotating_t} \, , \\
  A^{(0),a}_{\phi}&=\left( \chi_1 \sqrt{1-\chi_1^2} \sin (\phi -t  \omega ), \chi_1 \sqrt{1-\chi_1^2} \cos (\phi -t  \omega ), \left(\chi_1^2-1\right)\right)  )\label{eq:Ext_Gauge_Rotating_phi} \, ,
\end{align}
with the three $SU(2)$ chemical potentials $\mu^a$.
While the dual current is given as,
\begin{align}
  \braket{J_t^a} = A^{(2),a}_{t}&=\left(-\frac{\left(2 \chi _1^2 \chi _2-\chi _2\right) \omega  \sin (\phi -\tau  \omega )}{2 \sqrt{1-\chi _1^2}},\frac{\left(2 \chi _1^2 \chi _2-\chi _2\right) \omega  \cos (\phi -\tau  \omega )}{2 \sqrt{1-\chi _1^2}},-\chi _1 \chi _2 \omega  \right), \label{eq:ROT_current_t}\\
  \braket{J_\phi^a}=  A^{(2),a}_{\phi}&=\left(\frac{\left(2 \chi _1^2 \chi _2-\chi _2\right) \sin (\phi -\tau  \omega )}{2 \sqrt{1-\chi _1^2}},-\frac{\left(2 \chi _1^2 \chi _2-\chi _2\right) \cos (\phi -\tau  \omega )}{2 \sqrt{1-\chi _1^2}},\chi _1 \chi _2 \right) \label{eq:ROT_current_phi}\, .
\end{align}
Again we check the divergence of this current via $\partial_{i}\braket{J^{i,a}}$ which gives,
\begin{equation}\label{eq:divJRotating}
    \partial_{i}\braket{J^{i,a}}=\frac{ \left(2 \chi _1^2-1\right) \chi _2 \left(16 a^2-\omega ^2 \left(\sqrt{16 a^2-\omega^2 } +1\right)\right) }{32 a^2 \sqrt{1-\chi _1^2}}\left(\cos (\phi -t  \omega ), \sin (\phi -t \omega ),0 \right).
\end{equation}
A few observations are in order here as well. First, a simple interpretation of the non-conservation is due to the oscillating chemical potential inducing an oscillating current. However in this case we have additional dependence on the compact coordinate $\phi$. This leads to a chemical potential with the form of a traveling wave whose wave vector has the value $k=1$ and angular frequency is $\omega$. Accompanying the charge density ``wave'' propagating in the compact $\phi$ direction is a current wave $\vev{J^{a}_{\phi}}$. The current wave lags behind the charge wave by a phase of $\pi$ as can be seen by inspection of eq.\ (\ref{eq:ROT_current_t}) and eq.\ (\ref{eq:ROT_current_phi}). Furthermore the divergence of the current is also a propagating wave in the compact $\phi$ direction. This divergence of the current lags the charge wave by a phase of $\pi/2$. This can be seen in figure~\ref{fig:currentwaves}.
Meanwhile the charge density $\vev{J^{a}_t}$ and chemical potential $\mu^a$ are always in phase. By suitable adjustment of the parameters $\chi_1$ or $\chi_2$ we may have a conserved current $\vev{J}$ ($\chi_1=\pm 1/\sqrt{2}$). Interestingly if we take $\omega=0$ all sources $\mu^a$ and vacuum expectation values $\vev{J^a_t}$ vanish leaving an external gauge field $A^{a}_{\phi}$ along with a persistent current along the compact $\phi$ direction. Furthermore it is interesting to note that as in the static case although the four-current is non-conserved, the current averaged over one cycle of period $T=2\pi/\omega$ is conserved. 
Note also that the magnitudes of the chemical potential vectors $\mu^a$ and charges $\langle J_t^a\rangle$ which are three-vectors in gauge space, have magnitudes which are constant in time.

It should be noted that associated with the external $SU(2)$ gauge field in the dual field theory is a field strength which has non-zero components in this case. 
\begin{align}
    F^1_{t\phi}=-F^1_{\phi t}&= \chi _1 \sqrt{1-\chi _1^2} \omega  \cos (\phi -t  \omega ), \\
   F^2_{t\phi}=-F^2_{\phi t}&=   \chi _1 \sqrt{1-\chi _1^2} \omega  \sin (\phi -t  \omega )\, .
\end{align}

\textbf{Topological solution:} For the topologically non-trivial solution we have only an external gauge field in the dual theory. That is we have no bulk radial dependence of $A_{\mu}\exd x^{\mu}$. In this sense the gauge field in the bulk represents a trivial lifting of the gauge field in the boundary theory to the bulk theory
\begin{equation}\label{eq:liftOfA}
    L(A_{\mu})=\tilde{A}_{\mu}=(A_{i},0) \, ,
\end{equation}
where $\tilde{A}$ is the bulk five dimensional gauge field. It can be expanded near the $AdS$-boundary simply as,
\begin{equation}
    \tilde{A}_{\mu}=(A_{i},0)r^0 \, .
\end{equation}
In the notation of the previous section we have the following gauge field configuration in the boundary field theory given below broken into the three non-zero gauge field components. For the component in the $\psi$-direction, $A^{(0),a}_{\psi}$, we find,
\begin{subequations}
\begin{align}
   A^{(0),1}_{\psi}&=-n \cos (\theta )\, ,\\
   A^{(0),2}_{\psi}&=-n  \sin (\theta ) \cos (\phi )\, ,\\
    A^{(0),3}_{\psi}&=-n \sin (\theta ) \sin (\phi )\, .
 \end{align}\label{eq:Top_current_psi}
    \end{subequations}
For the component in the $\theta$-direction, $A^{(0),a}_{\theta} $, we find,
    \begin{subequations}
    \begin{align}
    A^{(0),1}_{\theta}&=\sin (n \psi )\sin (\theta )  \cos (n \psi ) \, ,\\
    A^{(0),2}_{\theta}&=\sin (n \psi )\left( \cos (\theta ) \cos (\phi ) \cos (n \psi )+\sin (\phi ) \sin (n \psi )\right) \, , \\
     A^{(0),3}_{\theta}&=\sin (n \psi ) \left(\cos (\phi ) \sin (n \psi )-\cos (\theta ) \sin (\phi ) \cos (n \psi )  \right)\, ,
     \end{align}\label{eq:Top_current_theta} 
    \end{subequations}
    and for the component in the $\phi$-direction, $A^{(0),a}_{\phi} $, we find,
    \begin{subequations}
    \begin{align}
    A^{(0),1}_{\phi}&= \sin ^2(\theta ) \sin ^2(n \psi )\, ,  \\
     A^{(0),2}_{\phi}&= \sin (\theta )\sin (n \psi )\left( \sin (\phi )  \cos (n \psi )- \cos (\theta ) \cos (\phi ) \sin (n \psi )\right)\, ,  \\
      A^{(0),3}_{\phi}&=- \sin (\theta ) \sin (n \psi )\left( \cos (\theta ) \sin (\phi ) \sin (n \psi )+\cos (\phi ) \cos (n \psi )\right) .
\end{align}\label{eq:Top_current_phi}
\end{subequations}
Notably the field strength associated with this gauge field configuration has non-zero $F_{\psi\theta},F_{\psi\phi},F_{\theta\phi}$, which act as angular magnetic fields along $\phi,\,\theta,\,\psi$, respectively. Due to the length of the expressions we do not display them here. While there is an external gauge potential and field strength in the dual theory, the vacuum expectation value of the current is zero for this configuration
\begin{equation}
    \langle J_{i}^{a} \rangle = A^{(2),a}_{i}= 0\, .
\end{equation}

We note that it is interesting that this lifted bulk gauge field sources the Einstein equations and solves both the Einstein-Skyrmion and Einstein-Yang-Mills equations of motion. It would be interesting to investigate if this type of solution can be generalized to solutions with non-vanishing expectation value of the dual field theory current.

\subsection{Energy-momentum tensor}
\label{sec:Energy_Momentum}
The holographic energy-momentum tensor can be calculated via standard techniques as laid out in~\cite{deHaro:2000vlm,Taylor:2000xw}. A space-time metric in the Fefferman-Graham coordinate system can be written in the form,
\begin{equation}
    ds^2=\frac{1}{4\rho^2}\exd\rho^2+\frac{g_{ij}(x,\rho)}{\rho}\exd x^{i}\exd x^{j},
\end{equation}
where $g_{i j}$ admits the following expansion,
\begin{equation}
    g=g_{(0)}+g_{(2)}\rho+g_{(4)}\rho^2+h_{(4)}\rho^2\log(\rho)+\cdots
\end{equation}
The appropriate holographic relation is~\cite{Fuini:2015hba},
\begin{equation}
    \vev{T_{ij}}=\frac{1}{\kb}\left(g_{(4)ij}-g_{(0)ij}\tr g_{(4)}-(\log(\Lambda) +\mathcal{C})h_{(4)ij}\right),\label{eq:holo_Energy_Momentum}
\end{equation}
where $\mathcal{C}$ is an arbitrary scheme dependent constant and $4\pi G_5=\kb$. We utilized eq.\ (\ref{eq:holo_Energy_Momentum}) to construct the quasi-local stress-energy tensor of Brown and York~\cite{Balasubramanian:1999re} for all solutions we presented in section~\ref{sec:blackBraneSolutions}. This stress-energy tensor has been shown~\cite{Skenderis:2008dg} to be dual to the vacuum expectation value of the energy-momentum tensor in the dual field theory. 

We can also use this stress energy tensor to define conserved charges of the solutions~\cite{Balasubramanian:1999re}, given a Killing vector $\xi$ and a time-like unit vector $u^{\mu}$ normal to a space-like hypersurface $\Sigma$ there is an associated conserved charge,
\begin{equation}
  Q =-\int_{\Sigma}\sqrt{\text{det}(g_{(0)})}\xi_{i}u_{j}\vev{T^{ij}}.\label{eq:Conserved_Quant}
\end{equation}
For example, there is a time-like Killing vector associated with time translation $\xi^t$, and the conserved charge is the mass of the space-time, 
\begin{equation}
    M =-\int_{\Sigma}\sqrt{\text{det}(g_{(0)})}\xi^t_{i}u_{j}\vev{T^{ij}}.\label{eq:Conserved_Charge_Mass}
\end{equation}

\textbf{Static Solution:} For the static black brane we find the same energy-momentum relation as in the case of the Schwarzschild black brane with,
\begin{equation}
   \kb \braket{T^{00}}=\epsilon, \hspace{2cm} \kb\braket{T^{i'j'}}=p \delta^{i'j'},
\end{equation}
where the pressure $p=\epsilon/3$ and $i'\text{\& }j'\in (1,2,3)$. The energy-momentum tensor is traceless, indicating we have a conformal fluid in the dual field theory. Here $\epsilon$ the energy density is given as $\epsilon=3m_s/4$.

\textbf{Rotating Solution:} For the rotating solution we find,
\begin{subequations}
\begin{align}
   \kb \braket{T^{00}}&=\frac{(4 a+\omega ) \left(r_0^4 \omega -4 c_1\right)}{16 a^2}, \\
  \kb  \braket{T^{\phi\phi}} &= \frac{(4 a-\omega ) \left(r_0^4 \omega -4 c_1\right)}{16 a^2}, \\
   \kb \braket{T^{0\phi}}&=\frac{\sqrt{4 a-\omega } \sqrt{4 a+\omega } \left(r_0^4 \omega -4 c_1\right)}{16 a^2}\,,
\end{align} \label{eq:rotating_ENMom_Tensor}
\end{subequations}
with $ \braket{T^{ij}}=0$, for $i=x,y$ the non-compactified spatial coordinates. An 
elucidating choice of the coefficient $c_1$ can be seen from eq.\ (\ref{eq:rotating_ENMom_Tensor}) when taking $c_1=r_0^4a$, yielding,
\begin{subequations}
\begin{align}
  \kb   \braket{T^{00}}&=r_0^4\frac{\omega^2 -16 a^2}{16 a^2}, \\
  \kb  \braket{T^{\phi\phi}} &=r_0^4 \frac{\omega^2 -16 a^2}{16 a^2}, \\
  \kb  \vev{T^{0\phi}}&=r_0^4\frac{\sqrt{4 a-\omega } \sqrt{4 a+\omega } \left( \omega -4 a\right)}{16 a^2},
\end{align} \label{eq:rotating_ENMom_Tensor_Units}
\end{subequations}
with all other components of the energy-momentum tensor vanishing.
One can see that this solution represents a conformal fluid when one computes the trace of the energy-momentum tensor using the metric of the boundary theory $\gamma_{\mu\nu}=\text{Diag}(-1,1,1,1)$ and $\gamma_{0\phi}=\gamma_{\phi 0}=\omega/\sqrt{16a^2-\omega^2}$. Curiously we see this energy-momentum tensor encodes zero pressure in both the non-compact field theory directions, while the boundary fluid has both a momentum current 
and a pressure along the compact $\phi$ direction. One can understand this behavior by considering a singly-spinning Myers-Perry (Kerr) black hole in $AdS_5$, i.e.~with the two angular momenta set equal to each other.
Interestingly, we see that our rotating fluid has a direct analogy to the energy-momentum tensor of the extremal Myers-Perry black hole~\cite{Myers:1986un,Hawking:1998kw,Hawking:1999dp,Gibbons:2004ai,Gibbons:2004js,Gibbons:2010cr}. 
In the conventions of~\cite{Garbiso:2020puw}, 
the energy density and pressures are given by
\begin{equation}
  \braket{T^{00}}=\frac{\mu\left({a_{\text{MP}}}^2+3L^2\right)}{8\pi G_5L^4}, \hspace{.5cm} 
 \braket{T^{11}}=\braket{T^{22}} = \frac{\mu\left(L^2-{a_{\text{MP}}}^2\right)}{8\pi G_5L^4},\hspace{.5cm}
    \braket{T^{33}}=\frac{\mu\left(L^2+3{a_{\text{MP}}}^2\right)}{8\pi G_5L^4}\label{eq:Meyers_Perry},
\end{equation}
where it should be noted that the Myers-Perry ${a_{\text{MP}}}$ is the angular momentum per unit mass. Despite strong formal similarity, this is different from our $a$, which appears as an integration constant, a boundary condition on the metric function $F(r)$, see eq.~\eqref{eq:Rotating_Skyrme_Solution}. 
In the extremal limit $a_{\text{MP}}\rightarrow L$, and we find that the transverse pressures vanish,
\begin{equation}
    \braket{T^{00}}=\frac{\mu\left(4L^2\right)}{8\pi G_5L^4}, \hspace{.5cm} 
    \braket{T^{11}}=\braket{T^{22}} =0,\hspace{.5cm}
    \braket{T^{33}}=\frac{\mu\left(4L^2\right)}{8\pi G_5L^4}\label{eq:Meyers_Perry_extremal}.
\end{equation}

Applying our formula given in eq.\ (\ref{eq:Conserved_Charge_Mass}) to the energy-momentum tensor of our rotating solution eq.\ (\ref{eq:rotating_ENMom_Tensor}) gives,
\begin{equation}
    M=\frac{8 a^2 \left(r_0^4 \omega -4 c_1\right)}{G_5 (4a-\omega)^2 (4 a+\omega )}V,
\end{equation}
with $V=\int \exd^2x$. 
We can also apply eq.\ (\ref{eq:Conserved_Quant}) to obtain the conserved angular momentum associated with the azimuthal Killing vector where we find,
\begin{equation}
    J=\frac{2 a \left(r_0^4 \omega -4 c_1\right)}{G_5 \sqrt{(4 a-\omega )^3 (4 a+\omega )}}V.
\end{equation}
With the angular momentum and the mass we can verify that  the Euler relation,
\begin{equation}
    M=ST+J\Omega_T
\end{equation}
is indeed satisfied for our rotating solution where we have used the thermodynamic angular velocity $\Omega_T$ given in eq.\ (\ref{eq:Ang_Vel_Thermodynamics}). Using the relations for $M$ and $J$ also allow us to find an expression for both the coefficient $a$ and $c_1$ in terms of $\omega,M$ and $J$,
\begin{subequations}
\begin{align}
    a&=\frac{ \omega }{4 \sqrt{1-(J/M)^2}}, \label{eq:ang_mom_a}\\
   c_1&= \frac{\omega}{4 M}  \left(\frac{2 G_5 J^2}{\sqrt{1-J^2/M^2}} \left(1-\sqrt{1-J^2/M^2}\right)+M r_0^4\right).
\end{align}
\end{subequations}
Both parameters $a$ and $c_1$ are depend upon the rotation parameter of a Skyrmion in the internal space and the angular momentum per unit mass in the physical $AdS$ space-time. \\
Notice that we also have the following relation,
\begin{equation}
   a=\frac{ \omega }{4 \sqrt{1-(1/\Omega_T)^2}}, \label{eq:ang_mom_a}\\
\end{equation}
which is consistent with the definition of $\Omega_T$ given in eq.\ (\ref{eq:Ang_Vel_Thermodynamics}).

\textbf{Topological solution: } For the topologically non-trivial solution we find the following energy-momentum tensor,
\begin{align}
  \kb\vev{T^{00}}&=\frac{3}{16} L^4 \left(\frac{4 \log (\Lambda )}{e^2}+L^2+4\, m_t\right), \\
 \kb \vev{T^{\psi\psi}}&=\frac{L^2 \left(e^2 \left(4\, m_t-3 L^2\right)+4 \log (\Lambda )+4\right)}{16 e^2 n^2}, \\
  \kb\vev{T^{\theta\theta}}&=\frac{L^2 \csc ^2(n \psi ) \left(e^2 \left(4\, m_t-3 L^2\right)+4 \log (\Lambda )+4\right)}{16 e^2}, \\
 \kb \vev{T^{\phi\phi}}&= \frac{L^2 \csc ^2(\theta ) \csc ^2(n \psi ) \left(e^2 \left(4\, m_t-3 L^2\right)+4 \log (\Lambda )+4\right)}{16 e^2}.\label{eq:Topo_Energy_Momentum}
\end{align}
Notably, this topologically non-trivial solution has finite energy density.  
Here $\Lambda$ is an energy scale associated with our renormalization procedure revealing a logarithmic dependence of the energy-momentum tensor on the cutoff scale\footnote{In our expression for the energy momentum tensor we have chosen the finite scheme dependent coefficient $\mathcal{C}$ of eq.\ (\ref{eq:holo_Energy_Momentum}) to take the value $\mathcal{C}=5/4$. This value has been chosen to cancel an explicit finite contribution to the energy density stemming from the gauge field configuration. This choice of $\mathcal{C}$ shifts the contribution of the gauge field from the energy density to the pressures.}.
This indicates that there should be a special solution at 
$|e|= e_c =\frac{2 \sqrt{\log (\Lambda )+1}}{\sqrt{3 L^2-4 m_t}}$
for which the pressures vanish. If $|e|<e_c$, then the pressures are negative. Despite first appearances, the existence of such a {\it potential} limitation does not invalidate the $e\to 0$ limits taken in the previous sections. In that limit, the renormalization scale has to be fixed to $\Lambda=\text{exp}(-1)$ in order to adequately represent physics at the scale of the system\footnote{We note that this choice of $\Lambda=\text{exp}(-1)$ is the smallest possible value of the renomalization scale. }. 
Tracing over the energy-momentum tensor with the near boundary metric given by,
\begin{equation}
   \exd s^2= \sin ^2(n \psi ) \left(\exd\theta ^2+\exd\phi^2 \sin ^2(\theta )\right)-\frac{\exd t^2}{L^2}+\exd\psi^2 n^2,
\end{equation}
reveals two trace contributions,
\begin{equation}
\kb \vev{\tensor{T}{^{i}_{i}}}=\frac{3 L^2}{4 e^2}-\frac{3 L^4}{4}. \label{eq:Trace_Anomaly}
\end{equation}
Each has its own origin.\footnote{Note that the first of these two trace contributions $3 L^2/(4 e^2)$ has nothing to do with either the conformal anomaly of the $\mathcal{N}=4$ SYM theory we consider here, nor with its R-charge current (chiral) anomaly. It arises from an explicit breaking of the conformal invariance by introducing the coupling scale $e$~\cite{Fuini:2015hba}. We use this occasion to recall that in our model the R-charge current (chiral) anomaly is absent since we chose to work at vanishing Chern-Simons coupling on the gravity side.}  
We note in passing that the trace vanishes by fixing $e=1/L^2$. 
We can trace the origin of the first piece to the presence of an external field strength $F^a_{ij}$.
It turns out that the first term in eq.~\eqref{eq:Trace_Anomaly} has nothing to do with the conformal anomaly of $\mathcal{N}=4$ SYM theory. Instead, it is well known to originate from the coupling to an external gauge field~\cite{Fuini:2015hba,Taylor:2000xw}. 
This coupling breaks the conformal symmetry explicitly, leading to a non-zero trace of the energy-momentum tensor. 
Hence, we can expect one trace contribution to the energy-momentum tensor to be proportional to the trace over the square of the field strength~\cite{Fuini:2015hba,Taylor:2000xw},
\begin{equation}\label{eq:externalFTrace}
    \kb \vev{\tensor{T}{^{i}_{i}}}\propto -\frac{1}{4}\tr \tilde{F}^2, \hspace{1cm} \tilde{F}=\frac{1}{3e}F \, .
\end{equation}
The second term in eq.~\eqref{eq:Trace_Anomaly} has a deeper field theory meaning and corresponds to the conformal anomaly of $\mathcal{N}=4$ SYM theory on the space-time $\mathbb{R}\times S^3$~\cite{Balasubramanian:1999re,Henningson:1998gx},
\begin{equation}
  -\lim_{\epsilon\rightarrow 0} \frac{L^4}{8\pi G_5}\left(-\frac{1}{8}R^{\mu\nu}R_{\mu\nu}+\frac{1}{24}R^2\right)=-\frac{3 L^4}{4\kb},
\end{equation}
where $z=1/r=\epsilon$ is a cutoff surface on which we construct the Ricci tensor $R_{\mu\nu}$ and Ricci scalar $R$. 
Finding the conformal anomaly value consistent with the direct field theory computation confirms the holographic relation of the Einstein-Skyrme theory to $\mathcal{N}=4$ SYM, as discussed at the beginning of this section.

Applying our formula for the conserved charge associated with time translations eq.\ (\ref{eq:Conserved_Charge_Mass}) to the energy-momentum tensor eq.\ (\ref{eq:Topo_Energy_Momentum}) we find,
\begin{equation}
    M=\frac{3 \pi  L^2 n}{32 G_5}+\frac{3 \pi\,  m_t\, n}{8 G_5}+\frac{3 \pi  n }{8 e^2 G_5}\log \left(\Lambda L\right) \, . \label{eq:Mass_equation}
\end{equation}
This mass contains three terms, the second term, for $n=1$, is the standard result for a global $AdS$ black hole ($M_{BH}$) in five dimensions~\cite{Horowitz:1998ha}. While the first term, again for $n=1$ is the mass of the global $AdS$ ($M_{AdS}$) space-time~\cite{Balasubramanian:1999re}. The third term is the result of the renormalization scale dependence of the energy-momentum tensor. What is interesting is that for a solution with winding number $\mathpzc{q}=n=1$ these results agree with the standard results without an $SU(2)$ gauge field up to a renormalization point dependence. While for $\mathpzc{q}=n>1$ we find,
\begin{equation}
   \mathring{M}\equiv M-\frac{3 \pi  n }{8 e^2 G_5}\log \left(\Lambda L\right)= M_{BH}(n)+M_{AdS}(n),
\end{equation} 
for $M_{BH}(n)=n M_{BH}(1)$ and $M_{AdS}(n)=n M_{AdS}(1)$. In this way the imprint of the Skyrmion field onto the energy of the space-time is clear. The masses of topological Skyrme-$AdS_5$ black holes can be measured in units of global $AdS$ and black hole masses up to a renormalization point mass.  

\subsection{Topological charge}
As discussed in section~\ref{sec:topologicalCharges},
 we will apply the definition of the topological charge~\eqref{eq:instantonCharge} to our solutions. 
Our solutions are pure gauge everywhere in space-time, hence,
\begin{equation}
 \mathpzc{q}
= - i \frac{e^3}{12\pi^2} \oint\limits_{S^3}\exd^3 x \, \hat n_i \epsilon^{ijmn}  \text{tr}(A_j A_m A_n) 
= \frac{1}{24\pi^2} \oint\limits_{S^3}\exd^3 x \, \hat n_i \epsilon^{ijmn}   \text{tr}\left( \partial_j \omega \omega^{-1} \partial_m \omega \omega^{-1} \partial_n \omega \omega^{-1} \right)
     \,  .
\end{equation}
We compute this quantity 
from our bulk gauge field solutions, evaluating them on the $S^3$ located at the boundary of our $AdS_5$ space-time. This topological charge is also the topological charge of the boundary field configuration due to eq.~\eqref{eq:liftOfA}. 
For both the static and rotating solutions the dual gauge fields given by eq.~\eqref{eq:Ext_Gauge_static} for the static case and by eq.~\eqref{eq:Ext_Gauge_Rotating_t} and eq.~\eqref{eq:Ext_Gauge_Rotating_phi} for the rotating case do not represent maps covering the internal $S^3$ an integer number of times~\cite{Ioannidou:2006nn}. 
This is because in both cases only one of the spatial components is non-zero. 
This is due to the fact that this ansatz restricts dependence to the coordinates $t$ and $r$. Derivatives acting in other directions vanish leading to $K_i=0$. The winding number $\mathpzc{q}$ is a topological quantity. Hence, if it vanishes in one coordinates system, then it vanishes in all. 

Therefore the black branes considered in Section 4.1 have zero topological charge. However the topologically nontrivial solutions presented in section~\ref{sec:blackBraneSolutions} carry a non-zero topological charge. |BH{ The use of capitals for the sections is inconsistent her} Computing the charge density we find,
\begin{equation}
i \epsilon^{ijmn} \text{tr}\left( \partial_j \omega \omega^{-1} \partial_m \omega \omega^{-1} \partial_n \omega \omega^{-1} \right)  =\left( \frac{3}{4}  n \sin (\theta ) \sin ^2(n \psi ),0,0,0\right),
\end{equation}
leaving us with\footnote{Note that we had suggestively chosen to label the integer parameter in the solution~\eqref{eq:Topo_non_trivial_Metric} by $n$.}
\begin{equation}
\mathpzc{q}=\frac{2}{3\pi^2} \int_0^{\pi}\exd\theta \int_0^{\pi}\exd\psi  \int_0^{2\pi}\exd\phi \frac{3}{4}  n \sin (\theta ) \sin ^2(n \psi )=n \in\mathbb{N}.
\end{equation}
In summary, the gauge configuration of the Schwarzschild black hole with Skyrmion hair has a dual gauge configuration which wraps the internal $S^3$ a total of $\mathpzc{q}=n$ times. As seen in the previous section, the mass of the space-time grows with each wrapping $n$ of the internal space-time.

\subsection{Comparison to the Sakai/Sugimoto model}
\label{sec:comparisoToSS}
In table~\ref{tab:gaugeGravitySkyrmions} we have already pointed out differences and similarities between the Sakai/Sugimoto model~\cite{Sakai:2004cn}, the Son/Stephanov model~\cite{Son:2003et}, and our model. Now that we have presented the details of our model, a few more comments are in order. 

Our setup is rather simple despite being a top-down model embedded in a pure gauge sector of type IIB supergravity. Consequently, it allows for analytic solutions at $T\neq 0$. Remarkably, our solutions take into account the backreaction of the gauge field configurations (equivalent to our bulk Skyrmions) on the geometry. Previous works  utilize the top-down Sakai/Sugimoto $D4/D8-\bar{D8}$-brane construction. There is a stack of $N$ D$4$-branes which generates an effective background metric, see~\cite{Sakai:2004cn,Kruczenski:2003uq}, similar to the stack of $N$ D$3$-branes in the original correspondence~\cite{Maldacena:1997re}. 
The $D8$- and $\bar{D8}$ are technically involved due to the square root of the determinant appearing in the Dirac-Born-Infeld actions for each probe D$p$-brane (e.g.~a D4-brane has $p=4$), 
\begin{equation} \label{eq:DBraneAction}
    S_{Dp} = -\tau_p \int d^{p+1}x \, e^{-\phi} \sqrt{\text{det}(P[G]+P[B]+2\pi \alpha' F)}  \, ,
\end{equation}
with the probe brane tension $\tau_p/g_s = (2\pi)^{-p} {\alpha'}^{-(p+1)/2}/g_s$ where $g_s$ is the string coupling constant. Further, there is the background scalar $\phi$, as well as the pullback $P[\dots]$ of the background metric $G$, the pullback of the antisymmetric background two-form $B$, the gauge field strength $F$ living on the D$p$-brane, and its coupling constant and the inverse string tension $\alpha'$~\cite{Sakai:2004cn,Hata:2007mb}. Solutions often need to be obtained numerically, especially at non-zero temperature and chemical potential~\cite{Aharony:2006da,Parnachev:2006dn,Peeters:2006iu,Nawa:2006gv}. Generally, this is a common feature of top-down D-brane constructions, another example being the D3/D7-brane system~\cite{Mateos:2006nu,Babington:2003vm} at non-zero temperature and chemical potential~\cite{Kobayashi:2006sb,Apreda:2005yz}, and in particular when computing perturbations on top of the branes, e.g.~meson spectra~\cite{Erdmenger:2007ja,Myers:2008cj}, and confinement
\cite{Horigome:2006xu,Nakamura:2006xk,Aharony:2006da}. For example, the complexity of the $D4/D8-\bar{D8}$-brane system forces the authors of~\cite{Hata:2007mb} to abandon the non-Abelian DBI action and work only with the ``leading order'' Yang-Mills and Chern-Simons terms to make the computation feasible.

By definition, the backreaction of these probe branes on the background geometry is not taken into account, neither is the backreaction of e.g.~topologically non-trivial configurations of the gauge field $F$. This can be seen when counting the factors of $1/{\alpha'}^2\propto N$, where N is the number of colors, and comparing to the background action which is of order $N^2$. For example, the effect of D$7$-probe branes is suppressed by a relative factor of $1/N$. 

Going a step beyond probe branes, the backreaction of the $D8$-branes including simple gauge field configurations onto the geometry has been taken into account~\cite{Bigazzi:2014qsa}. However, topologically non-trivial gauge field configurations have not been backreacted as far as we know, and hence are a new feature of our solutions. 

The Skyrmion in the $D4/D8-\bar{D8}$ system emerges in the boundary field theory. As usual in quantum field theory, this boundary theory is defined in a fixed flat metric, i.e.~with the metric being an external field. 
These Skyrmions are identified as baryons. They are constructed from D4-branes wrapped on a non-trivial four-cycle in the background metric. 
Such a D4-brane is realized as a small instanton configuration in the world-volume gauge theory on the probe D8-brane. 
In the simplified Yang-Mills-Chern-Simons action~\cite{Hata:2007mb}, the effective boundary theory action 
is identical to the Skyrme model, in which baryons appear as solitons, referred to as Skyrmions, as we reviewed in section~\ref{sec:Skyrmions}. 
The baryon number of such a Skyrmion, as we know, is defined as the winding number carried by the pion field. This winding number is the instanton number in the five-dimensional Yang-Mills theory. 

It is remarkable that within the $D4/D8-\bar{D8}$ model the authors~\cite{Sakai:2004cn, Hata:2007mb} consider the three spatial boundary coordinates and the radial $AdS$ coordinate as the Euclidean space on which the topological properties are evaluated. This is also true for the Son and Stephanov~\cite{Son:2003et}. As reviewed in section~\ref{sec:topologicalCharges}, in a higher dimensional (larger than four) space-time, any four-dimensional subspace may be considered in order to assess the topological properties of e.g.~gauge field configurations. We have chosen to Wick rotate the time coordinate and consider topological quantities in the four-dimensional subspace spanned by Euclidean time and the three spatial boundary directions. This is another point that sets our model apart from previous considerations. 

Our model also allows phases and transitions among them which differ from those discussed already in the literature. The Sakai/Sugimoto model allows two distinct brane constructions, and a transition between them at a critical temperature, $T_c$. This is the confinement/deconfinement transition in a dual gauge theory~\cite{Witten:1998zw}. In the low temperature (confined) phase at $T<T_c$ chiral symmetry is broken, while it is restored in the high temperature (deconfined) phase at $T>T_c$~\cite{Aharony:2006da,Parnachev:2006dn}. Our topological Skyrme-$AdS_5$ black hole solution realizes the confinement/deconfinement transition as it undergoes a variation on the Hawking-Page transition, as discussed above. In addition, since our gauge field configuration or equivalently the Skyrmion is backreacted on the geometry, there are topologically distinct phases carrying distinct Chern numbers which are accessible in our model. These may be realized depending on their free energy~\eqref{eq:DeltaFHawkingPage}. This also begs the question if topological transistions are realized in the Sakai/Sugimoto model, e.g.~between QCD vacua. Our model may serve as a construction tool in this regard. 

Our Skyrme model deconfinement transition temperature depends on the topological charge $\mathpzc{q}$, which is a label of the dual field theory state, and the Skyrme coupling $e$. This means, in our model, one (holographic QCD) vacuum with topological charge $\mathpzc{q}$ undergoes a deconfinement transition at a temperature distinct from that of a vacuum with a distinct $\mathpzc{q}$. Our model may be useful to qualitatively study mechanisms for QCD vacua to transition into each other. 
Note that recently in the Sakai-Sugimoto model 
transitions between false holographic vacuum states have been considered~\cite{Bigazzi:2020phm}. There the rate for producing true vacuum bubbles in a metastable phase has been computed. This has been done utilizing the deconfinement transition and the chiral symmetry breaking transition. Similar considerations in our model would be interesting, allowing vacuum bubbles with distinct gauge field topology.

It would be interesting to relate our solutions and phases to those considered by Sakai and Sugimoto. Their D4/D8-brane model~\cite{Sakai:2004cn,Hata:2007mb} is constructed in type IIA superstring theory. Our model may be considered to be embedded in type IIB superstring theory. 
Type IIA is related to type IIB theory by T-duality. If one of the 10 coordinates in either theory is compactified onto an $S^1$ with a certain radius $R$, strings can wrap around this $S^1$ a number of $W$ times. There is also a momentum along that direction, $K$. T-duality exchanges $W$ and $K$, between type IIA theory on a small compactification radius, to IIB theory on a large compactification radius, and vice versa. 
One is tempted to speculate if T-duality could help to relate our Skyrme solutions to those of Sakai/Sugimoto. 

More directly, we are able to 
show 
at least strong similarity between our solution and those discussed in~\cite{Hata:2007mb}. For this purpose we take a black brane limit of our black hole solution. We utilize a stereographic projection in order to unwrap our compact geometry $S^3\to \mathbb{R}^3$,
\begin{equation}
\left(X,Y,Z\right)=\frac{\sin (n \psi )}{1-\cos (n \psi )} \left(\cos (\theta ),\sin (\theta ) \cos (\phi ),\sin (\theta ) \sin (\phi )\right) \, ,
\end{equation}
applied to the $4D$ metric,
\begin{equation}
  \exd s^2=  \frac{4 \left(\text{dX}^2+\text{dY}^2+\text{dZ}^2\right)}{\left(X^2+Y^2+Z^2+1\right)^2}-\frac{1}{L^2}\text{dt}^2 \, .
\end{equation}
Now using the scaling transformation $X^i\rightarrow \alpha x^i/2$, $x^i=(x_1,x_2,x_3)$, $t\rightarrow\alpha t$ and series expanding around $\alpha=0$, we find that the leading order contribution is at $O(\alpha^2)$,
\begin{equation}
    \exd s^2\sim -\frac{1}{L^2}\exd t^2+\delta_{ij}\exd x^i\exd x^j \, .
\end{equation}
In order to compare, take $g$ from eq.~(3.3) in~\cite{Hata:2007mb} and Taylor expand it around $(z,X1,X2,X3)=\vec{0}$, with $Z=1$,
\begin{equation}
 g\sim  \frac{ 1}{\sqrt{x_1^2+x_2^2+x_3^2+1}} \left(
\begin{array}{cc}
 -1-i x_3 & -x_2-i x_1 \\
 x_2-i x_1& -1+i x_3\\
\end{array}
\right) \,.\label{eq:HataSolutionForComparison}
\end{equation}
Now take our $SU(2)$ group element $U$ and use the coordinate transformation above and series expand around $\alpha=0$ and take the leading order contribution in $\alpha$,
\begin{equation}
 \left(
\begin{array}{cc}
 -1+i \alpha  x_3 & \alpha  x_2+i \alpha  x_1 \\
 -\alpha  x_2+i \alpha  x_1 & -1-i \alpha  x_3 \\
\end{array}
\right) \, .
\end{equation}
Finally, inverting all coordinates, $(x_1,x_2,x_3)\to (-x_1,-x_2,-x_3)$, gives,
\begin{equation}\label{eq:ourUForComparison}
    \left(
\begin{array}{cc}
 -1-i \alpha  x_3 & \,\, -\alpha  x_2-i \alpha  x_1 \\
 \alpha  x_2-i \alpha  x_1 & \, \,  -1+i \alpha  x_3 \\
\end{array}
\right) \, .
\end{equation}
Now we squint our eyes a bit, remove the counting parameter by setting $\alpha=1$, and ignore overall factors, attributing them to the brane embedding. Then the expanded group element eq.~\eqref{eq:ourUForComparison} looks remarkably 
similar to the group element  eq.~\eqref{eq:HataSolutionForComparison}, as we had advertised. Differences may be due to the distinct metrics in which these gauge fields live. 

Note again that we have considered different sets of coordinates in the two models. In our model, we choose to work with the three spatial boundary coordinates, $(x_1,x_2,x_3)$, and the time, $t$. In contrast to that, from~\cite{Hata:2007mb} we see the authors chose three spatial boundary coordinates and the radial $AdS$ coordinate, $z$.
As reviewed in section~\ref{sec:topologicalCharges}, both choices are valid, and it is interesting that in our comparison they lead to gauge field configurations strikingly similar in structure.

In application to QCD the Sakai/Sugimoto model has been very successful in predicting observables and matching experimental results. It would be interesting to investigate if our simpler solutions could be predictive in a similar fashion. Potentially, they may help in gaining a simplified but analytic understanding of the underlying physics.

Finally, we put our topological solutions from section~\ref{sec:topologicalMeronSolutions} into the context of known realizations of baryons within the Sakai-Sugimoto model, reviewed e.g.~in~\cite{Suganuma:2020jng}. 
The standard realization of the baryon in QCD 
is in the form of
the soliton solutions, referred to as Skyrmions in (3+1) dimensions. 
A second realization is the instanton in holographic QCD~\cite{Sakai:2004cn}. 
Recently~\cite{Suganuma:2020jng}, it has been shown that the Abrikosov vortex in (1+2)-dimensional Abelian Higgs theory~\cite{Witten:1976ck} 
is the topological equivalent of 
the instanton in holographic QCD. In this sense, there is a realization of the baryon in Abelian Higgs theory.  
Merons have recently been considered as topological solutions, baryons, within holographic QCD~\cite{Suganuma:2020jng}. In their construction, which is distinct from ours, the authors find that a single meron carries infinite energy. Hence, they construct a two-meron solution of finite size and with finite energy. 
In contrast to that, our topologically non-trivial solutions are not localized in the boundary spatial coordinates $\{x,\,y,\,z\}$. Instead, they are uniform throughout boundary space and thus have no spatial size. Our solutions are also uniform along the $AdS$ radial coordinate $r$, which can be seen from eq.~\eqref{eq:unit} as the gauge field configuration does not depend on $r$. 
In that sense, we have a family of non-local gauge field configurations, labeled by second Chern numbers. 
Furthermore, our gauge field configurations have finite energy density. This can be seen from the energy-momentum tensor~\eqref{eq:Topo_Energy_Momentum} which contains the contributions from the gauge field configuration and the geometry (loosely speaking the ``mass of the black hole''), see eq.~\eqref{eq:Mass_equation}.

In addition to the baryon realizations discussed above, there is the baryon from brane-induced Skyrmions~\cite{Nawa:2006gv}. These are finite size baryons when truncated to light mesons~\cite{Hata:2007mb}. However, they are pointlike when such a truncation is not performed~\cite{Hata:2007mb}. 
In that case, including a Chern-Simons term brings it back to a finite size, curing the infinite gradient problem of a pointlike configuration. 
Our model fits into this picture as a distinct analytic holographic realization of a gauge field configuration with topological charge $\mathpzc{q}=1$, which may be interpreted as baryon number. Our solutions take into account the full backreaction of the non-Abelian gauge field onto the geometry. These are solutions within a (4+1)-dimensional Einstein-Skyrme theory. This is currently excluding a possible Chern-Simons-term, but introducing such a term is possible.\footnote{In fact, the Chern-Simons term with a specific coupling strength is required for this theory to be dual to $\mathcal{N}=4$ Super-Yang-Mills theory.} Just like the finite baryon size effects found to result from Chern-Simons contributions in the Sakai/Sugimoto model, this may have interesting consequences for our solutions. 

In this work, we utilize the $SU(2)$-Einstein-Yang-Mills formulation of the Einstein-Skyrme theory in order to obtain a holographic interpretation. It would be interesting to understand a direct relation between the Skyrmion in the bulk and the observables in the boundary theory. This may reveal a more direct relation between our bulk Skyrmions and known representations of baryons in quantum field theories.

\section{Discussion \& application of results}
\label{sec:Discussion}
We have found new black brane/hole solutions which can either be regarded as solutions to Einstein-Skyrme theory, or equivalently as meron solutions in Einstein-Yang-Mills theory, see section~\ref{sec:merons}. On one hand, in the context of the gauge/gravity correspondence, the meron point of view is more clear, see section~\ref{sec:holoSkyrmionsMerons}. The solutions in the field theory are states with a non-trivial $SU(2)$ gauge field configuration sourcing an $SU(2)$ current. On the other hand, the Skyrmion point of view may serve as a construction tool for new, non-trivial solutions, see section~\ref{sec:Skyrmions} and section~\ref{sec:blackBraneSolutions}. 
Two top-down embeddings into type IIB superstring theory are provided for the Einstein-Skyrme action~\eqref{eq:skyrmeAction} via its Einstein-Yang-Mills equivalent~\eqref{eq:meronAction}, see section~\ref{sec:holoSkyrmionsMerons}. The first, more straight forward interpretation of the $SU(2)$ symmetry is as a subgroup of the $R$-symmetry in the $\mathcal{N}=4$ Super-Yang-Mills theory. 
The second proposed embedding into a $D3/D7$-brane system is to be taken in the spirit of a heuristic motivation.\footnote{Clearly, our Einstein-Skyrme action lacks the relevant factors coming from the metric induced on the branes, see the brane action~\eqref{eq:DBraneAction}.}

Collecting the results of this work, we choose to separate them into two categories: gravitational results and field theory results. 
\paragraph{Gravity results:} 
We have found three types of analytic solutions to the Einstein-Skyrmion theory defined by the action~\eqref{eq:skyrmeAction}: 
\begin{enumerate}
    \item The {\it Static Skyrme-$AdS_5$ black brane}, eq.~\eqref{eq:SkyrmionSol} and~\eqref{eq:SkyrmionSolA}, is static in space-time and rotating in internal space, i.e.~rotating on the $SU(2)$ gauge manifold at frequency $\omega$. The blackening factor takes a form similar to that of an $AdS_5$ Reissner-Nordstr\"om black brane. The ``charge term'' in the blackening factor~\eqref{eq:SkyrmionSolA} contains a charge-like parameter, $\chi_2\omega$ determined by the internal rotation frequency. However, this parameter can not take any extremal value as there is only one horizon. The Hawking temperature $T$ in figure~\ref{fig:temperature}, and entropy density $s$, eq.~\eqref{eq:sStatic}, of this solution are determined by internal angular frequency $\omega$, mass $m_s$, and solution parameter $\chi_2$.\footnote{For simplicity, we mostly consider solutions without the kinetic Skyrmion term, i.e.~at vanishing pion coupling, $f_\pi=0$. However, we have shown that these are smoothly connected to solutions with non-zero pion coupling $f_\pi\neq 0$, eq.~\eqref{eq:blackening_factor_kinetic}.}
    \item The {\it Rotating Skyrme-$AdS_5$ black brane}, eq.~\eqref{eq:rotatingBraneSoln}, is rotating in space-time with relative angular frequency $\Omega_T$. This solution shares various properties with extremal $AdS_5$ Myers-Perry (Kerr) black holes. It has vanishing temperature, $T=0$, but a non-zero entropy density $s$~\eqref{eq:sRotating}. 
    One interesting aspect of the holographic interpretation of the Einstein-Skyrme model is that it reduces the arbitrariness of the Skyrmion solution in the rotating case.
    The Einstein-Skyrme equations of motion for the rotating black brane system have an additional freedom due to the hidden symmetry associated with the imposition of the constraint. This leaves either the Skyrme field or a metric component free. We choose to fix the metric and leave the matter contribution free. However this freedom is removed when considering the mapping of the Einstein-Skyrme system to an Einstein Yang-Mills theory. This leads to an additional set of equations (the curved space Yang-Mills equations) which must also be satisfied. This additional set of equations in our system reduces to a single equation which fixes the solution of the Skyrme field. 
    \item The {\it Topological Skyrme-$AdS_5$ black holes}, eq.~\eqref{eq:Topo_non_trivial_Metric}, is a family of solutions asymptoting to global $AdS_5$, i.e.~with a compact horizon and boundary field theory topology $S^3\times \mathbb{R}$.
    Here, the Skyrmion solution has a non-zero winding number $\mathpzc{q}=n\in \mathbb{N}$, parametrizing this family of solutions. In other words, this solution wraps the spatial $S^3$ $n$ times around the $S^3$ representing the $SU(2)$ group manifold. 
    Again, while the solution is known analytically, the horizon location as a function of the black hole mass and the Hawking temperature are calculated numerically and displayed in figure~\ref{fig:Topologically_Non_Trivial}. For a Skyrmion coupling constant, $e\gg 1$, the horizon and temperature behave similarly to that of a global $AdS_5$ black hole. In that case there is a minimum temperature. However, when the Skyrmion becomes strong, $e\to 0$, the temperature develops an additional maximum at a small value for the radius. 
    Furthermore, this solution undergoes a Hawking-Page transition at a transition\footnote{
    The Hawking-Page phase transition has been studied for rotating M2, M5, and D3-branes~\cite{Cai:1998ji}. Since in our framework we found rotating planar black branes and also topological black holes undergoing the Hawking-Page transition, it may be interesting to compare to phase transitions in~\cite{Cai:1998ji}.
    }
    temperature altered by the Skyrmion dynamics and defined by the zeros of the free energy difference~\eqref{eq:DeltaFHawkingPage}. 
    The free energy depends on both the Skyrme coupling $e$ and the winding number, $n$, of the gauge configuration, see figure~\ref{fig:Topologically_Non_Trivial_Free_Energy}. 
\end{enumerate}
All three solutions are equivalent to Einstein-Yang-Mills (meron) solutions. 
%

\paragraph{Field theory results:} 
The holographically dual field theory is $\mathcal{N}=4$ coupled to an $SU(2)$ (non-)conserved current sourced/driven by the three oscillating $SU(2)$ chemical potentials. These chemical potentials are the non-normalizable mode coefficients given in eq.\ (\ref{eq:Ext_Gauge_static})
for the static Skyrme-$AdS_5$ black brane, in
eq.~\eqref{eq:Ext_Gauge_Rotating_phi} for the rotating Skyrme-$AdS_5$ black brane, and
eq.~\eqref{eq:Top_current_phi}, \eqref{eq:Top_current_psi}, \eqref{eq:Top_current_theta} for the topological Skyrme-$AdS_5$ black hole. 
They act as sources for the $SU(2)$ current $J^\mu$ in the quantum field theory. 
The field theory states dual to each of the three solutions have the following properties: 
\begin{enumerate}
    \item Static Skyrme-$AdS_5$ black brane. 
    The chemical potentials $\mu^a$ for the static solution are  oscillating with frequency $\omega$ in internal space as time progresses. As a consequence, the divergence of the current $J$ is periodically oscillating around zero, eq.~\eqref{eq:divJStatic}. Hence the $SU(2)$-charge is conserved on average but not at any given instant of time. 
    The $SU(2)$-charges $J^{\mu,a}$ and the spatial current components $J^{\mu,a}$ in $a=1,2$ gauge directions are oscillating over time with frequency $\omega$, eq.~\eqref{eq:vevStatic}.
    In this state, the energy and pressure are identical to the state dual to the Schwarzschild solution. 
    \item Rotating Skyrme-$AdS_5$ black brane. 
   For this solution, as for the static solution, the chemical potentials are oscillating. However, there is an additional spatial source in the $\phi$-direction, which is the source for the component $J^{\phi,a}$. This component  oscillates in internal space, and in one of the spatial angle directions as time progresses, both with frequency $\omega$. The divergence of the current $J$ oscillates periodically, and the time-averaged $SU(2)$-charge is conserved (see eq.\ (\ref{eq:divJRotating})). 
 The dual $SU(2)$ current components in the $a=1,2$ gauge directions are oscillating with frequency $\omega$. Expressions for all current components are given by eq.~\eqref{eq:ROT_current_t} and \eqref{eq:ROT_current_phi}. 
    A hallmark of this solution is its vanishing pressure in the two spatial directions perpendicular to the axis of rotation (see the energy-momentum tensor~\eqref{eq:rotating_ENMom_Tensor_Units}). This behavior is reminiscent of the extremal rotating Myers-Perry (Kerr) $AdS_5$ solutions. Angular momentum is flowing along one of the angle directions, namely $\phi$. That current, the energy density and the pressure in the $\phi$-direction depend on the frequency $\omega$ and an ``angular momentum'' parameter $a$ given in eq.\ (\ref{eq:ang_mom_a}). 
    \item Topological Skyrme-$AdS_5$ black hole.
    For the topological Skyrme-$AdS_5$ black hole, the sources for the spatial currents are spatially modulated. 
    All of the components of the dual $SU(2)$ current vanish in this state and its divergence is trivially zero. 
    In contrast to the static black brane, the chemical potentials $\mu^a =0$. 
    The topological solution has an energy density which depends only on the black hole mass $m_t$, eq.~\eqref{eq:Topo_Energy_Momentum}. However, its three pressures are equal to each other and grow with $m_t$ but decrease when $e\to 0$. These states stand out because they carry non-zero topological charge $\mathpzc{q}=1,\,2,\,3, \dots$, and hence the conserved mass and the entropy of the space-time are quantized. 
    The conformal anomaly, including one contribution from the Euler density external gauge field, is correctly reproduced by our gravity calculation. This is a further confirmation that our interpretation of this setup as $\mathcal{N}=4$ SYM plus a conserved current on $S^3\times\mathbb{R}$ is correct.
\end{enumerate}

\paragraph{Application to QCD and heavy-ion collisions:} 
As an application of our analytically-known topologically non-trivial state, one may consider the QCD phase diagram. Our state undergoes a deconfinement transition and can be used to model the effect of topological (e.g.~isospin or flavor) gauge field configurations.  
Remarkably, in our model the gauge field coupling $e\sim \gamma$ (of e.g.~isospin or flavor) and the topological charge $\mathpzc{q}$ determine the critical temperature of the deconfinement transition. The effect of the field topology on deconfinement depends on the coupling constant $e$. 
In particular, the critical point of the phase diagram may be studied in extensions of this model. 
Is it possible to generate topology-changing deconfinement transitions? Or is it possible to change the topology of a state above or below the deconfinement transition as the temperature changes?  
Also, could such a topology change occur along the renormalization group scale (which is dual to the radial $AdS$ coordinate)? 
A recent holographic model allows a transition between large and small black holes and also models the deconfinement transition~\cite{Arefeva:2020byn}. Such a transition may also occur within our model.  
Our rotating states may also contribute to the understanding of vortical quark-gluon-plasma states~\cite{STAR:2017ckg}. 
Such applications of our model will teach us at least qualitative lessons and yield proposals for similar effects in QCD. 

As is true for most holographic models, the Einstein-Skyrme model is very general in its predictions. It does not allow quantitative predictions, e.g.~for heavy-ion experiment results. However, this drawback is also a strength of holography: Our model describes and highlights general (topological and/or rotation) features of strongly coupled quantum many body systems. Hence, it is applicable to a large class of systems. Next, we exploit this generality to discuss a different field to which our results can be applied.

\paragraph{Applications to condensed matter topological phases:} 
Topological phases and topological phase transitions in condensed matter systems, such as fractional quantum Hall systems, constitute another field of application of our analytic solutions. In particular our non-Abelian gauge field dynamics may prove useful to manipulate quantum states which are stable against thermal and quantum fluctuations. 
The topological solutions we found may allow the transition of a field theory state with a particular winding number $\mathpzc{q}$ to a state with a different winding $\mathpzc{q}'$ dynamically. This constitutes a change in topology, i.e. a topological phase transition. Could such a change occur along the renormalization group scale (which is dual to the radial $AdS$ coordinate)? 
It would also be interesting to understand what kind of transition the Hawking-Page transition would model in a condensed matter system. 

In our model states contain charge density waves. They occur as responses to oscillating chemical potentials. Another effect is an oscillating angular current wave $J^\phi\sim \text{sin}(\phi-\omega \tau)$.  
These charge density waves are determined by the Skyrme coupling parameter $e$.

\paragraph{Outlook:} 
It would be interesting to give the bulk Skyrmion field $U$ in eq.\ (\ref{eq:top_Ansatz}) a dependence on the radial $AdS_5$ position $r$. For this purpose one may consider hypersurfaces perpendicular to the radial direction. The $AdS$-boundary is one such hypersurface. From eq.~\eqref{eq:instantonCharge} we see that then the topological charge on such a hypersurface is given by $n = n(r_{\text{cutoff}})$. This suggests the possibility of changing the topology of a given field theory  state as a function of the renormalization group scale (to which the radial $AdS$ coordinate is dual). 
Another important extension of our analysis here is to include a Chern-Simons term for the meron field. After working out how to encode this in the Einstein-Skyrme action, one would check if our solutions still solve the equations of motion which now contain Chern-Simons contributions, or if other analytic solutions result from the inclusion of a Chern-Simons term. 

Relations between topology and entanglement are fundamentally important to an understanding of multi-particle quantum states. Topological entanglement~\cite{Kitaev:2005dm,Levin:2006zz} and quantum entanglement~\cite{Padmanabhan:2020frr,Kauffman2002QuantumEA} have been discussed. Holographic realization was provided for example in~\cite{Pakman:2008ui,Nishioka:2009un}. As an example, ~the entanglement entropy in the (3+1)-dimensional toric code~\cite{3DToricCode}, identifies topological order at non-zero temperature below a critical temperature. However, \cite{NussinovOrtiz2008} disagrees, stating that the quantum entanglement is a non-local property of a quantum system; topological entanglement is a non-local property of the gauge manifold. 
An analytic understanding of the underlying fundamental physics of the interplay between entanglement and topology may be obtained from the (3+1)-dimensional field theory states found in this paper, as well as (2+1)-dimensional states~\cite{Harms:2019cag}.  

It would further be interesting to put our system into a magnetic field, splitting physical phenomena such as hydrodynamic transport effects into a subsystem perpendicular to the magnetic field and one longitudinal to it,  see e.g.~\cite{Ammon:2017ded,Ammon:2020}. This would reveal the influence of topology on the transport effects. 

On the technical side, one may want to answer the question of whether or not a near-extremal solution can be found, e.g.~by perturbing the components of the metric and/or the meron/Skyrmion. Similarly, 
it would be interesting to know if all three of our solutions are smoothly connected to solutions with an infinitesimal kinetic term (small $f_\pi$). 
One could examine our extremal rotating solution further to determine its physical properties, and to find non-extremal solutions in order to model vortical fluids. 
In \cite{Heusler:1991xx,Heusler:1993ci} the linear stability of self-gravitating Skyrmions against small, time-dependent perturbations was investigated for the particle-like solutions of the Einstein-Skyrme equations on a four-dimensional static background.  
The dynamical stability of the solutions we obtained in the five-dimensional version of the Skyrme model coupled to gravity have not been investigated in this paper.  
An investigation of the stability of the static, the rotating black brane, and the topologically non-trivial global $AdS$ black hole solutions is another interesting future project.

Finally, our model could serve as a basis for revealing new connections between high energy phenomena and condensed matter phenomena. Our model QCD deconfinement transition with its topological labels suggests there exists a counterpart in topological condensed matter materials. What does this transition imply for e.g.~strongly correlated electron systems, quantum Hall states, or for topological insulators? Our rotating Skyrme-$AdS_5$ solution may provide a model for a rotating highly vortical quark-gluon-plasma. What is the condensed matter analog of such a state? Can topological phases and topological phase transitions known from condensed matter physics lead to a better understanding of the QCD vacuua and transitions between them? What can be learned about toplogical phases and transitions in condensed matter materials from QCD vacuum transitions and topological gluon field configurations? Our model provides a testing ground for these questions.

\section*{Acknowledgements}
This research was supported in part by DOE grant DE-SC-0012447. We are grateful to A.~Cotrone and S.~Sugimoto for valuable comments on the draft of this work. MK thanks R.~Thomale for valuable discussions.

\appendix

\section{Considering the constraint equation} 
\label{app:constraintEquation}
We now turn to a discussion of the constraint~\eqref{eq:constraint}.  
The rotating Skyrme-$AdS_5$ black brane solution was obtained by imposing a constraint between metric tensor elements.  The constraint minimizes the matter contribution to the Lagrangian density and reduces the number of independent field equations, which is a characteristic of a hidden symmetry of the action.  One future goal will be to determine the effect of this hidden symmetry on the conformal field theories which live on the four-dimensional boundaries of the five-dimensional space-times which can be generated by various choices of the undetermined fields. 
A second goal could be to further develop the correspondence between the Skyrmion model parameters and the expected corrections to the field theory correlators. For now, we state our observations and suspicions. 

One observation is that eq.~\eqref{eq:constraint} is satisfied for the rotating Skyrmion solution described in section~\ref{sec:rotatingBBs}, but it is not satisfied for the static black brane in section~\ref{sec:staticBBs}. Furthermore, we have shown that the constraint~\eqref{eq:constraint} is satisfied at the horizon of an extremal Myers-Perry solution~\cite{Myers:1986un}.~\footnote{The Myers-Perry black hole is a solution with the three possible angular momenta equal to each other which is asymptotic to $AdS_5$. Extremality implies that the black hole has reached its maximum possible angular momentum. At this extremal charge value, the inner and outer horizon coincide while the temperature vanishes.} 
We found the rotating Skyrmion solution to have features similar to extremal Myers-Perry black holes, namely vanishing temperature, vanishing transverse pressure, but a non-zero angular momentum. Hence, it seems plausible that the constraint~\eqref{eq:constraint} is related to extremality. 
It would be interesting to test this suspicion either in other extremal black hole examples or to find a general derivation. 

Another question is if there is a symmetry associated with the constraint~\eqref{eq:constraint}. In this respect, we note that charged extremal black hole solutions in supergravity theories can exhibit supersymmetry enhancement at the horizon. These are called BPS black holes because they preserve part of the supersymmetry by saturating the Bogomol'nyi-Prasad-Sommerfield (BPS) bound. An example is the extremal dyonic Reissner-Nordstr\"om solution of $\mathcal{N}=2$ supergravity in 3+1 dimensions~\cite{Gibbons:1982fy}. 
The extremal BTZ black hole (in $AdS_3$) preserves some supersymmetry~\cite{Coussaert:1993jp,Hawking:1998kw}, and hence is a BPS solution. The mass of the extremal BTZ saturates the BPS bound with angular momentum $|J| = M/\ell$ with the $AdS_3$ radius $\ell$. Furthermore, it was found that such supersymmetric $AdS_3$ black holes, as well as their $AdS_5$ cousins, have to rotate, otherwise there is no regular horizon~\cite{Banados:1992gq, Gutowski:2004ez}. The rotating BPS black holes in $AdS_5$ can be oxidized to solutions of type IIB supergravity in $AdS_5\times S^5$~\cite{Gutowski:2004yv}. 

These examples and observations suggest that our five-dimensional rotating solution --when embedded into $AdS_5\times S^5$-- may be an extremal solution which potentially preserves part of the type IIB supergravity supersymmetry, i.e.~a BPS solution.\footnote{We note that the Skyrmion action vanishes when evaluated on-shell at our rotating solution~\eqref{eq:rotatingBraneSoln}.} 
Interpreting our solution as a meron solution to an Einstein-Yang-Mills theory implies that the dual field theory would preserve part of the $\mathcal{N}=4$ supersymmetry. 
It is further interesting to note that generally in the context of solutions to partial differential equations the BPS bound yields a series of inequalities. These  depend on the homotopy class of the solution at infinity. This set of inequalities is useful for solving soliton equations.  In some cases, the original equations can be reduced to the simpler Bogomol’nyi equations. This may be why the constraint~\eqref{eq:constraint}, if it is related to the BPS bound, simplifies the Einstein-Skyrmion equations. Scrutinizing these suspicions is beyond the scope of this paper and constitutes a task for future work.

\bibliographystyle{JHEP}
\bibliography{geo_pub}
\end{document}